\documentclass[pra,twocolumn,showkeywords,longbibliography,superscriptaddress,10pt，]{revtex4-1}
\usepackage{appendix}
\usepackage{graphicx}
\usepackage{dcolumn}
\usepackage{color}
\usepackage{times}
\usepackage{bm}
\usepackage{float}

\usepackage{amssymb}
\usepackage{amsmath}
\usepackage{epsfig}
\usepackage{epstopdf}
\usepackage{comment}
\usepackage{dsfont}
\usepackage{subfigure}
\usepackage{tikz}
\usepackage[colorlinks, citecolor=blue, linkcolor=blue,urlcolor=blue]{hyperref}
\usepackage[mathscr]{euscript}
\makeatletter
\renewcommand{\maketag@@@}[1]{\hbox{\m@th\normalsize\normalfont#1}}
\makeatother
\begin{document}
	\renewcommand{\thefootnote}{\fnsymbol {footnote}}
	
	 \title{Quantum coherence and entanglement in wireless quantum batteries}
	%
	\author{Jun-Jie Jiang}
	\affiliation{State Key Laboratory of Opto-Electronic Information Acquisition and Protection Technology, School of Physics, Anhui University, Hefei
		230601,  People's Republic of China}
	
	
	\author{Jin-Di Cao}
	\affiliation{State Key Laboratory of Opto-Electronic Information Acquisition and Protection Technology, School of Physics, Anhui University, Hefei 230601,  People's Republic of China}
	
	\author{Yu-Fei Zhang}
	\affiliation{State Key Laboratory of Opto-Electronic Information Acquisition and Protection Technology, School of Physics, Anhui University, Hefei 230601,  People's Republic of China}

    \author{Meng-Long Song}
	\affiliation{State Key Laboratory of Opto-Electronic Information Acquisition and Protection Technology, School of Physics, Anhui University, Hefei 230601,  People's Republic of China}
	
	\author{Dong Wang} \email{dwang@ahu.edu.cn}
	\affiliation{State Key Laboratory of Opto-Electronic Information Acquisition and Protection Technology, School of Physics, Anhui University, Hefei
		230601,  People's Republic of China}

	\date{\today}
	
	\begin{abstract}

We investigate the charging dynamics and thermodynamic performance of a wireless quantum battery system mediated by a common structured bosonic environment. By employing a unified resource-theoretic analysis, we elucidate the distinct roles of non-Markovian memory effects and coupling symmetry in regulating energy transfer. In the Markovian weak-coupling regime, we identify a transformative mechanism where the dynamic reconstruction of $l_1$-norm coherence compensates for the monotonic decay of first-order coherence to sustain energy transport. Conversely, the non-Markovian strong-coupling regime facilitates a cooperative resonance, characterized by the synchronized oscillation of entanglement and stored energy induced by environmental backflow. Furthermore, we reveal that coupling symmetry acts as a critical control parameter: while asymmetric coupling favoring the battery optimizes energy gain in memoryless environments, symmetric coupling under strong interactions unlocks a dark-state protection mechanism, effectively trapping energy within a decoherence-free subspace. Finally, a thermodynamic analysis based on ergotropy demonstrates that first-order coherence establishes the activation threshold for incoherent work, whereas $l_1$-norm coherence serves as the explicit fuel for coherent work extraction. These findings provide a refined theoretical framework for engineering environment-assisted quantum energy storage devices.
	\end{abstract}
	
	\maketitle

	\section{Introduction}
    {
    Quantum batteries (QBs), capable of storing energy within discrete quantum degrees of freedom, offer a promising paradigm for efficient energy capture and transfer \cite{nine,ten}. Unlike their classical counterparts, QBs exploit quantum phenomena to achieve superlinear charging powers and superior work extraction capabilities \cite{11,12,13,14,15,16,17,18}. However, a fundamental challenge remains: any realistic quantum system is inevitably open, interacting with its surrounding environment \cite{one}. Such interactions typically induce decoherence and dissipation, degrading the very quantum properties—such as coherence and entanglement—that sustain the battery's performance \cite{one}. Consequently, developing strategies to preserve these resources against environmental noise is critical for the realization of robust quantum energy storage devices \cite{two,three,four,five,six,seven,eight,sxq1,sxq3}. Promising approaches include Floquet engineering to reactivate dissipative batteries \cite{Bai2020} and schemes to mitigate self-discharging \cite{Song2025}, which effectively suppress the degradation of stored energy.
    
    Quantum resources serve as the fuel for these advanced capabilities. Quantum entanglement, arising from non-local correlations, allows multiple battery cells to charge cooperatively, providing the theoretical foundation for contactless, long-range wireless charging protocols \cite{two,30a}. Notably, recent advances have demonstrated that remote charging protocols can be engineered to suppress degradation, offering new perspectives for wireless energy transfer \cite{Song2024}.Simultaneously, quantum coherence is identified as a prerequisite for coherent extractable work, ensuring that stored energy can be converted into useful thermodynamic tasks \cite{one,31a,12}. Together, these resources not only surpass classical limitations in power and efficiency but also open pathways for next-generation energy systems with high capacity and aging resistance \cite{two,33a}. Very recently, the opportunities and challenges regarding the thermodynamic limits and experimental implementations of quantum batteries have been comprehensively reviewed \cite{Ferraro2026}.

 Recently, significant progress has been made in optimizing QB performance through specific physical models and mechanisms. For instance, Dou et al. investigated the role of many-body interactions in various systems, such as the cavity Heisenberg spin chain \cite{dfq1}, the extended Dicke model \cite{dfq2}, and the Rosen-Zener model \cite{dfq3}. They demonstrated that atomic interactions and quantum phase transitions can substantially enhance charging power and stored energy. Furthermore, they proposed a robust charging scheme based on superconducting transmon qubits \cite{dfq4}, utilizing dark-state mechanisms to suppress decoherence. In the context of environment-mediated energy transfer, Song et al. explored the dynamics of open QBs sharing a common reservoir. Their works revealed that the tightness of entropic uncertainty relations \cite{ml1,ml2} and Einstein-Podolsky-Rosen (EPR) steering \cite{ml3} can serve as effective indicators for energy transfer efficiency, highlighting the beneficial role of non-Markovian memory effects in sustaining wireless charging.

     Although these studies have provided valuable insights into environmental effects and many-body dynamics, the specific dynamics of wireless charging mediated by a common environment---where the charger and battery share the same dissipative reservoir---remain under-explored \cite{four,35a,22,sxq2}. {In particular, the utilization of non-reciprocity to achieve infinite energy storage in quantum batteries during wireless dissipative charging mediated by a common environment—a remarkable performance—has stimulated much interest in wireless charging protocols \cite{nrQB2,prapplied2025}. Under this charging framework, monitoring changes in energy-dominated battery performance and exploring the reasons why multiple resources synergistically promote energy growth represents an important topic \cite{qn3}.  In standard quantum battery architectures, since direct coherent coupling requires precise spatial proximity or dedicated physical wires between the charger and the battery, which imposes strict constraints on scalability and introduces unavoidable short-range cross-talk. In contrast, the environment-mediated wireless charging protocol offers the following advantages: (i) By embedding the components in a common bosonic reservoir (such as an optical cavity or a shared engineered bath), charging can be achieved remotely without any direct physical contact, bypassing spatial constraints.
(ii) Unlike direct coupling where energy continuously oscillates back and forth and eventually leaks away in the presence of global dissipation, the shared reservoir enables a cooperative dark-state protection mechanism under symmetric coupling. This effectively traps the stored energy within a decoherence-free subspace, drastically mitigating self-discharging.
(iii) While environmental coupling is traditionally viewed solely as a source of destructive decoherence, our protocol demonstrates that non-Markovian memory effects can be harnessed to dynamically reconstruct coherence and generate stable quantum channels.}
    
    In this paper, we investigate the charging dynamics and thermodynamic performance of a wireless quantum battery mediated by a common structured bosonic environment. We explicitly explore how non-Markovian memory effects and coupling configuration (symmetry) regulate the evolution of quantum resources. Our results demonstrate that non-Markovian strong coupling facilitates a cooperative resonance where entanglement and energy oscillate synchronously, whereas the Markovian weak coupling regime relies on a compensatory reconstruction of $l_1$-norm coherence to sustain energy transfer. Furthermore, we reveal that coupling symmetry acts as a crucial control parameter: asymmetric coupling enhances charging speed in dissipative limits, while symmetric coupling unlocks a dark-state protection mechanism for energy trapping. Finally, through a thermodynamic analysis based on ergotropy, we clarify the roles of different coherence measures, identifying first-order coherence as the activation threshold for incoherent work and $l_1$-norm coherence as the fuel for coherent work extraction. 
    
    The remainder of this paper is organized as follows. In Sec. II, we introduce the physical model of the wireless charging system and derive the exact analytical dynamics under both Markovian and non-Markovian regimes. Sec. III presents a detailed analysis of the charging performance and resource evolution, highlighting the distinct mechanisms governed by coupling symmetry and environmental memory. In Sec. IV, we perform a thermodynamic analysis using ergotropy to quantify the useful work and establish the relationship between quantum coherence and extractable energy. Finally, Sec. V summarizes our main findings and provides an outlook for future research.

 \section{MODEL AND THEORETICAL FRAMEWORK}

\subsection{Hamiltonian and Charging Setup}

{We focus on a wireless charging model consisting of three subsystems shown in Fig. \ref{f1}: a quantum charger ($A$), a quantum battery ($B$), and a common environment ($E$) that mediates the energy transfer, that is, there is no direct coupling term between the battery and the charger. }Both the charger and the battery are modeled as two-level systems with an identical transition frequency $\omega_0$ ($\omega_A = \omega_B = \omega_0$). The total Hamiltonian of the system, under the Rotating Wave Approximation (RWA) and setting $\hbar = 1$, is given by \cite{23, 24}:
\begin{equation}
    H = H_{0} + H_{E} + f(t)H_{I},
\end{equation}
where $H_0$ and $H_E$ describe the free energy of the qubits and the environment, respectively:
\begin{equation}
    H_{0} = \sum_{j=\{A,B\}} \omega_{0}\sigma_{j}^{+}\sigma_{j}^{-}, \quad H_{E} = \sum_{k}\omega_{k}a_{k}^{\dagger}a_{k}.
\end{equation}
Here, $\sigma_{j}^{+}$ and $\sigma_{j}^{-}$ are the raising and lowering operators for the $j$-th qubit ($j=A,B$), and $a_{k}^{\dagger}$ ($a_{k}$) are the creation (annihilation) operators for the $k$-th field mode with frequency $\omega_k$.

The interaction Hamiltonian $H_I$ describes the coupling between the qubits and the environmental modes:
\begin{equation}
    H_{I} = \sum_{k} \left[ g_{k}\mu_{1}(\sigma_{A}^{+}a_{k} + \sigma_{A}^{-}a_{k}^{\dagger}) + g_{k}\mu_{2}(\sigma_{B}^{+}a_{k} + \sigma_{B}^{-}a_{k}^{\dagger}) \right].
\end{equation}
In this expression, $g_k$ is the coupling constant for the $k$-th mode. The dimensionless real parameters $\mu_1$ and $\mu_2$ allow for the adjustment of the coupling strength for the charger and battery, respectively. We define the total coupling strength $\mu_T = \sqrt{\mu_1^2 + \mu_2^2}$ and the relative coupling weights $\xi_i = \mu_i / \mu_T$ ($i=1,2$). The switching function $f(t)$ controls the interaction timing: $f(t)=1$ for $t \in [0, \tau)$, and $f(t)=0$ otherwise.

\subsection{Spectral Density and Non-Markovian Dynamics}

The common environment is assumed to be a structured bosonic bath (e.g., a cavity field) characterized by a Lorentzian spectral density \cite{21}:
\begin{equation}
    J(\omega) = \frac{\xi^2 \lambda}{\pi [(\omega - \omega_0)^2 + \lambda^2]},
\end{equation}
where $\lambda$ is the spectral width, inversely related to the correlation time of the environment ($\tau_c \sim \lambda^{-1}$), and $\xi$ represents the effective coupling strength related to the vacuum Rabi frequency $R = \xi \mu_T$. The system enters the non-Markovian strong coupling regime when $R \gg \lambda$, where the memory effects of the environment become significant.

\subsection{Exact Dynamics}

Restricting our analysis to the single-excitation subspace, the time evolution of the probability amplitudes for the charger ($v_1$) and the battery ($v_2$) can be derived analytically. The amplitudes at time $t$ are given by \cite{23, 24}:
\begin{equation}
\begin{aligned}
    v_{1}(t) &= [\xi_{2}^{2}+\xi_{1}^{2}\kappa(t)]v_{01} - \xi_{1}\xi_{2}[1-\kappa(t)]v_{02}, \\
    v_{2}(t) &= -\xi_{1}\xi_{2}[1-\kappa(t)]v_{01} + [\xi_{1}^{2}+\xi_{2}^{2}\kappa(t)]v_{02}.
\end{aligned}
\end{equation}
Here, $v_{01}$ and $v_{02}$ denote the initial probability amplitudes for the charger and battery, respectively. The function $\kappa(t)$ characterizes the decoherence dynamics induced by the Lorentzian environment:
\begin{equation}
    \kappa(t) = e^{-\lambda t/2} \left[ \cosh\left(\frac{\chi t}{2}\right) + \frac{\lambda}{\chi} \sinh\left(\frac{\chi t}{2}\right) \right],
\end{equation}
where $\chi = \sqrt{\lambda^2 - 4R^2}$. The nature of $\chi$ determines the dynamical regime: a real $\chi$ corresponds to the weak-coupling Markovian regime, while an imaginary $\chi$ signifies the strong-coupling non-Markovian regime exhibiting oscillatory behavior.
    



 \begin{figure}
		\begin{minipage}{0.45\textwidth}
			\centering
			\subfigure{\includegraphics[width=8cm]{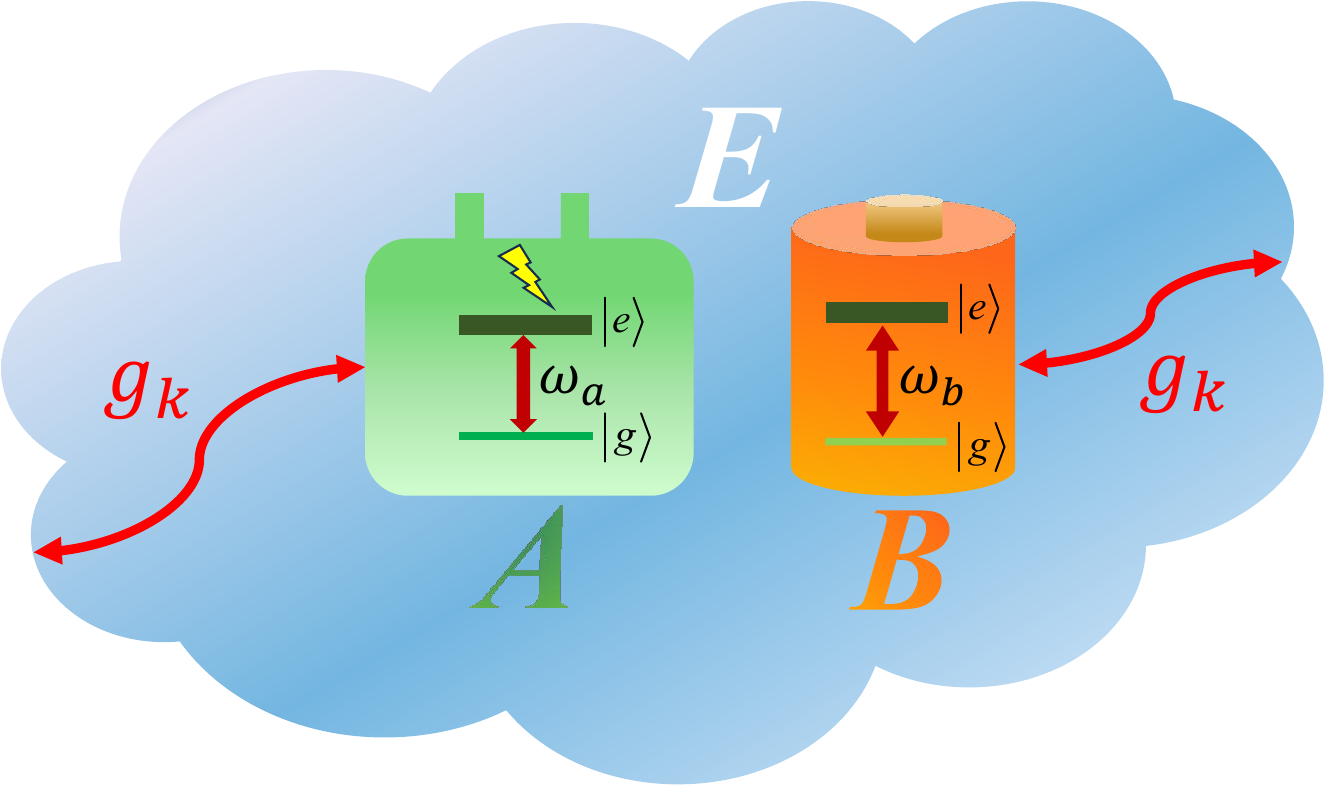}}
		\end{minipage}\hfill
		\caption{Schematic illustration of the wireless quantum charging model. The system consists of two two-level systems: a quantum charger $A$ (green sphere) and a quantum battery $B$ (orange sphere), characterized by their transition frequencies $\omega_a$ and $\omega_b$, respectively. We consider the resonant condition where $\omega_a = \omega_b = \omega_0$. They are spatially separated and embedded within a common dissipative environment $E$ (blue region). The red arrows indicate the interaction with the environmental field modes, characterized by the mode-dependent coupling strength $g_k$, which is further scaled by the dimensionless weights $\mu_1$ (charger) and $\mu_2$ (battery). Crucially, there is no direct coupling between $A$ and $B$; the energy transfer is mediated entirely by the indirect interaction through the common reservoir.}
		\label{f1}
	\end{figure}

\begin{figure}
		\begin{minipage}{0.5\textwidth}
			\centering
		\subfigure{\includegraphics[width=4.3cm]{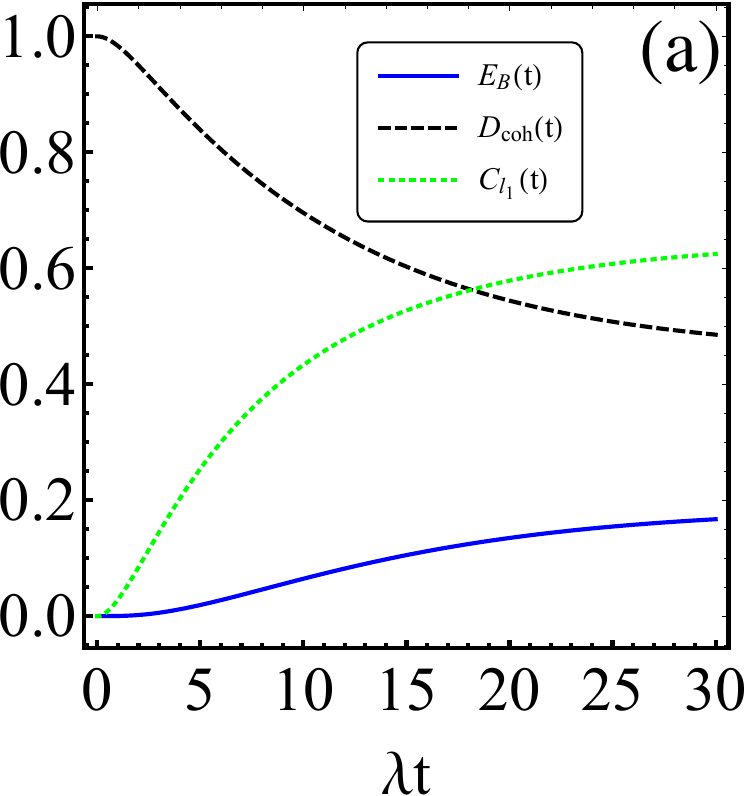}}
		\hspace{0.1cm}      
		\subfigure{\includegraphics[width=4.3cm]{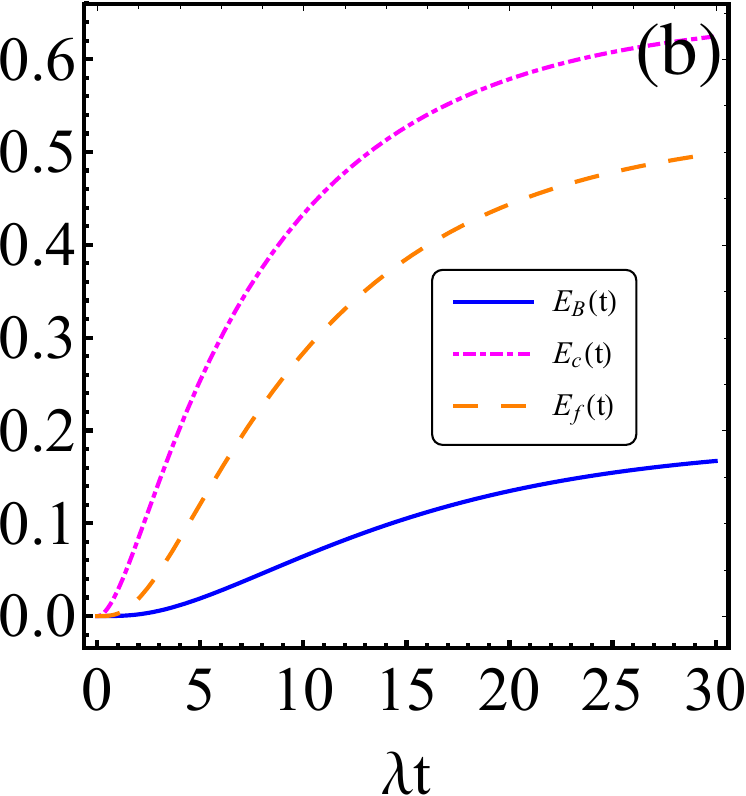}}
		\end{minipage}\hfill
		\caption{(Color online) Dynamics of quantum resources and energy storage in the Markovian weak-coupling regime ($R=0.3\lambda$). (a) Evolution of coherence: First-order coherence $D_{coh}$ (black dashed curve) decays monotonically, while $l_1$-norm coherence $C_{l_1}$ (green dotted curve) is reconstructed. (b) Evolution of entanglement and energy: Concurrence $E_C$ (pink dot-dashed curve) and Entanglement of Formation $E_f$ (orange long-dashed curve) precede the accumulation of battery energy $E_B$ (blue solid curve). The parameters are set to $\xi_1=0.5$ (charger coupling), $\xi_2=\sqrt{3}/2$ (battery coupling), corresponding to the battery-preferred configuration.}
		\label{f2}
	\end{figure}	
    \begin{figure}
		\begin{minipage}{0.5\textwidth}
			\centering
		\subfigure{\includegraphics[width=4.3cm]{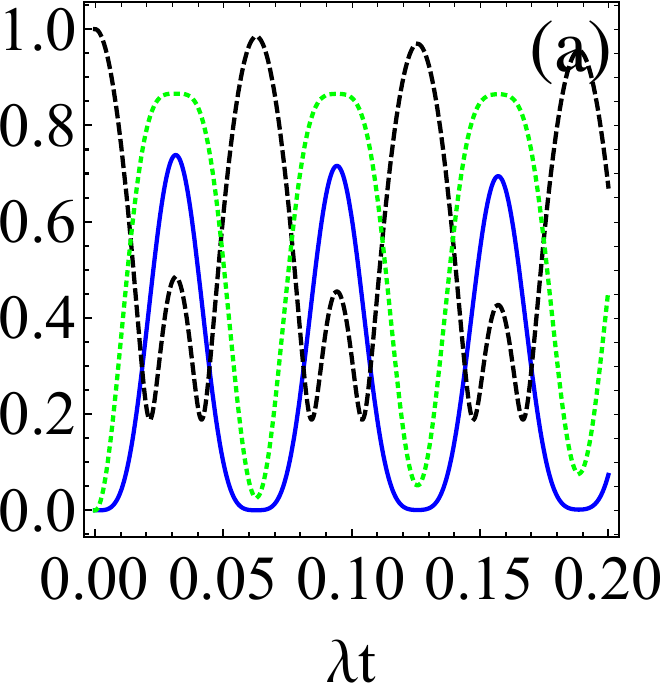}}
		\hspace{0.1cm}       
		\subfigure{\includegraphics[width=4.3cm]{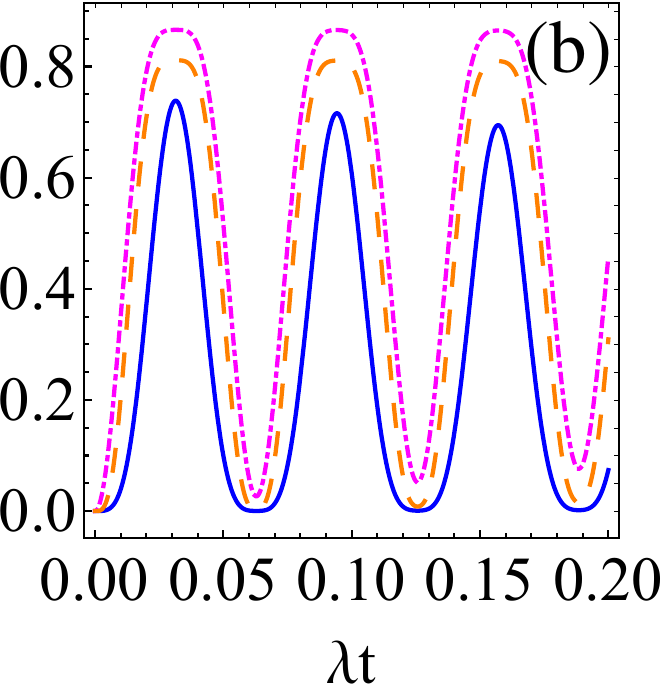}}
		\end{minipage}\hfill
		\caption{(Color online) Dynamics in the Non-Markovian strong-coupling regime ($R=100\lambda$). (a) Coherence protection: Both first-order coherence $D_{coh}$ (black dashed curve) and $l_1$-norm coherence $C_{l_1}$ (green dotted curve) exhibit persistent periodic oscillations, avoiding monotonic decay. (b) Resource-energy synchronization: Concurrence $E_C$ (pink dot-dashed curve), Entanglement of Formation $E_f$ (orange long-dashed curve), and battery energy $E_B$ (blue solid curve) oscillate in perfect phase. The parameters are $\xi_1=0.5, \xi_2=\sqrt{3}/2$ (battery-preferred coupling) with $\theta=\pi/2, \alpha=0$.}
		\label{f3}
	\end{figure}

\section{RESULTS AND DISCUSSION}

In this section, we investigate the charging performance and the evolution of quantum resources based on the exact dynamics derived in Sec.~II.

\subsection{Initial State Preparation}

To explore the interplay between energy transfer and quantum correlations, we consider a general initial state where the total excitation is shared between the charger and the battery, following the theoretical model described in Refs.~\cite{23, 24}. The system is initialized in the state:
\begin{equation}
    |\Psi(0)\rangle = \sin\theta e^{i\alpha} |e\rangle_A |g\rangle_B + \cos\theta |g\rangle_A |e\rangle_B.
\end{equation}
Here, the coefficient $v_{01} = \sin\theta e^{i\alpha}$ corresponds to the charger's initial excitation, while $v_{02} = \cos\theta$ represents the battery's state.

For the numerical analysis presented in this paper, we adopt the standard wireless charging configuration by fixing the parameters to $\theta = \pi/2$ and $\alpha = 0$. This corresponds to the scenario where the charger is initially fully excited and the battery is empty ($|\Psi(0)\rangle = |e\rangle_A |g\rangle_B$), allowing us to strictly evaluate the efficiency of environment-mediated energy transfer.

The dynamics are governed by the interplay between the spectral width $\lambda$ and the effective coupling strength $R$. In the following, we scale the time evolution by $\lambda t$ and consider different coupling configurations defined by the relative weights $\xi_1$ and $\xi_2$.


\subsection{Quantum Observables}

To quantify the charging performance and the evolution of quantum resources, we derive the reduced density operators for the charger and the battery from the total system state. Tracing out the environmental degrees of freedom and the other qubit, the reduced density matrices at time $t$ are diagonal in the energy basis $\{|e\rangle, |g\rangle\}$:
\begin{equation}
    \rho_{A}(t) = |v_{1}(t)|^{2}|e\rangle_{A}\langle e| + [1-|v_{1}(t)|^{2}]|g\rangle_{A}\langle g|,
\end{equation}
\begin{equation}
    \rho_{B}(t) = |v_{2}(t)|^{2}|e\rangle_{B}\langle e| + [1-|v_{2}(t)|^{2}]|g\rangle_{B}\langle g|.
\end{equation}
Based on these expressions, the stored energy in the system is directly given by its excited-state population relative to the transition frequency $\omega_0$:
\begin{equation}
 E_{A}(t) = \text{Tr}[H_{A}\rho_{A}(t)] = \omega_{0}|v_{1}(t)|^{2}.
 \end{equation}
   
\begin{equation}
    E_{B}(t) = \text{Tr}[H_{B}\rho_{B}(t)] = \omega_{0}|v_{2}(t)|^{2}.
\end{equation}
\begin{equation}
    \Delta E_{i} =E_{i}(t)-E_{i}(0), i={A,B}.
\end{equation}
Additionally, we monitor the quantum resources using multiple indicators to capture the rich dynamical features. We employ the $l_1$-norm of coherence ($C_{l_1}$) \cite{105} and the Entanglement of Formation ($E_f$), which is quantified via the concurrence \cite{100,101}. Furthermore, we analyze the first-order coherence $D_{coh}$ (off-diagonal elements), which corresponds to the standard optical visibility defined in quantum optics \cite{103}. The interplay between these measures provides a comprehensive view of the resource geometry \cite{102} and the complementarity between coherence and entanglement \cite{104}.

   \subsection{Markovian Regime with Battery-Preferred Coupling}

We first analyze the scenario where the system is weakly coupled to the environment ($R=0.3\lambda$), corresponding to the Markovian regime where memory effects are negligible. We consider an asymmetric coupling configuration favoring the battery ($\xi_2 > \xi_1$), specifically setting $\xi_1 = 0.5$ and $\xi_2 = \sqrt{3}/2$. The dynamical evolution is presented in Fig. \ref{f2}.

As shown in Fig. \ref{f2}(a), the evolution of quantum coherence reveals a competition between dissipation and environment-induced correlation. The first-order coherence $D_{coh}$ (black dashed curve) starts at unity and exhibits a monotonic decay. This behavior signifies the irreversible loss of the initial phase information due to the memoryless nature of the Markovian bath. In stark contrast, the $l_1$-norm coherence $C_{l_1}$ (green dotted curve) rapidly emerges from zero and stabilizes at a steady value. This phenomenon indicates a mechanism of \textit{coherence reconstruction}: while the intrinsic superposition of the charger is destroyed, the environment-mediated interaction generates new coherent components in the system basis, which are essential for sustaining the energy transfer process.

The interplay between entanglement and energy storage is depicted in Fig. \ref{f2}(b). Both the concurrence $E_C$ (pink dot-dashed curve) and the entanglement of formation $E_f$ (orange long-dashed curve) increase rapidly from zero, reaching their maxima significantly earlier than the battery's stored energy $E_B$ (blue solid curve). This temporal lead suggests that the establishment of quantum entanglement acts as a \textit{precursor} for energy storage . In this weak-coupling limit, the environment mediates the generation of entanglement, creating a quantum channel that subsequently facilitates the flow of energy. However, due to the continuous dissipation into the Markovian bath, the stored energy grows slowly and saturates at a relatively low level ($\Delta E_B < |\Delta E_A|$), confirming that entropy production in the memoryless environment limits the charging efficiency.

\begin{figure}[htbp]
    \centering
    \subfigure{\includegraphics[width=4.2cm]{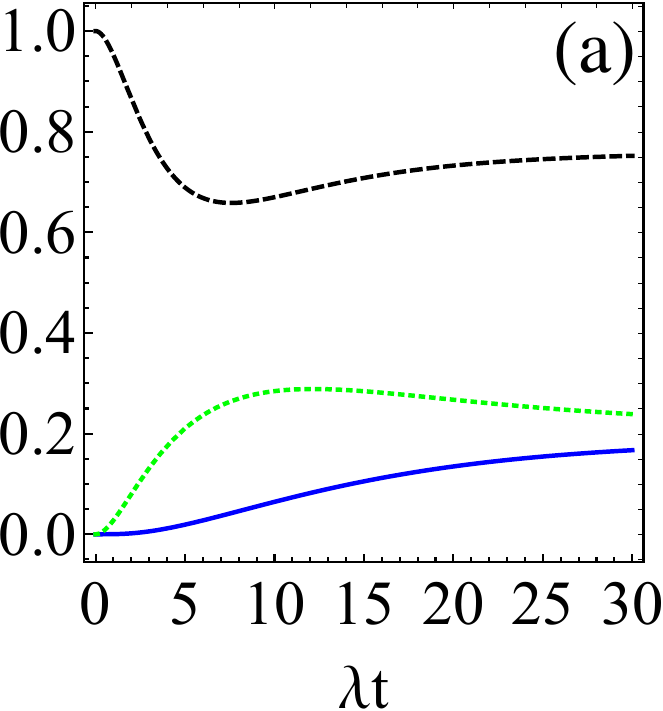}}
    \hspace{0.1cm} 
    \subfigure{\includegraphics[width=4.2cm]{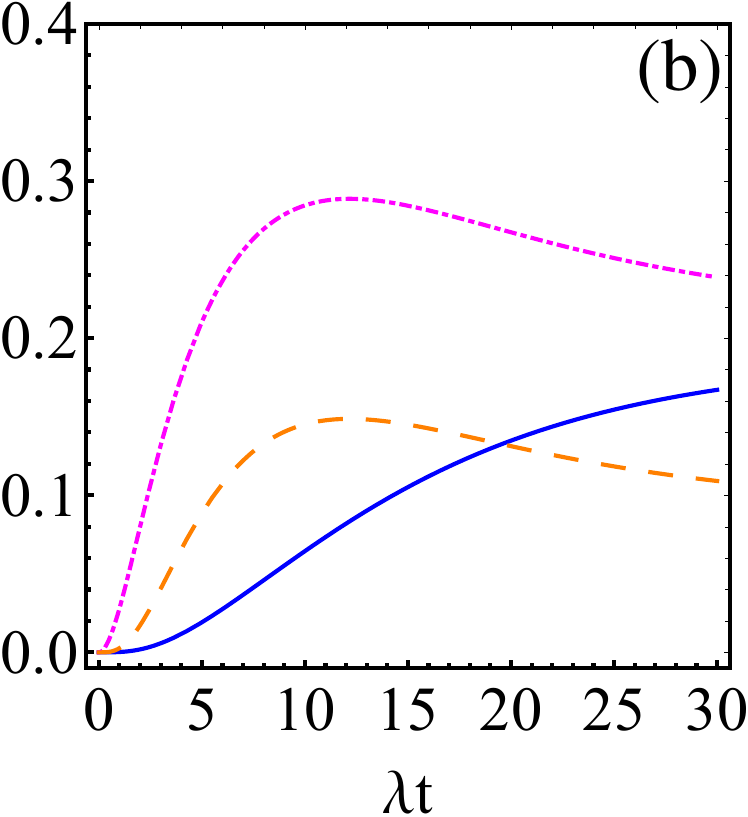}}
    
    \caption{(Color online) Dynamics in the Markovian regime with charger-preferred coupling ($R=0.3\lambda$). (a) Accelerated decoherence: First-order coherence $D_{coh}$ (black dashed curve) decays rapidly, and the reconstruction of $C_{l_1}$ (green dotted curve) is suppressed compared to Fig.~2. (b) Energy suppression: Entanglement (pink dot-dashed curve/orange long-dashed curve) is transient, and the battery energy $E_B$ (blue solid curve) is almost negligible. The parameters are $\xi_1=\sqrt{3}/2$ (stronger charger coupling) and $\xi_2=0.5$.}
    \label{f4}
\end{figure}

\begin{figure}[htbp]
    \centering
    \subfigure{\includegraphics[width=4.2cm]{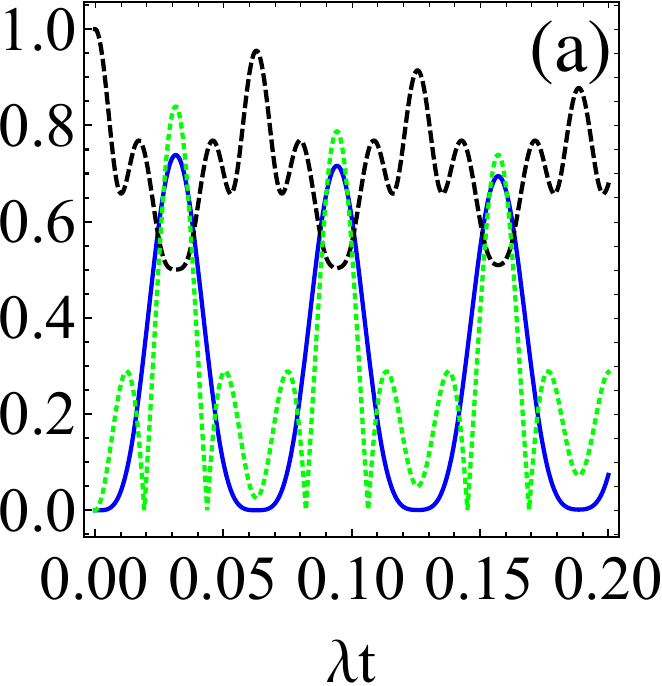}}
    \hspace{0.1cm}
    \subfigure{\includegraphics[width=4.2cm]{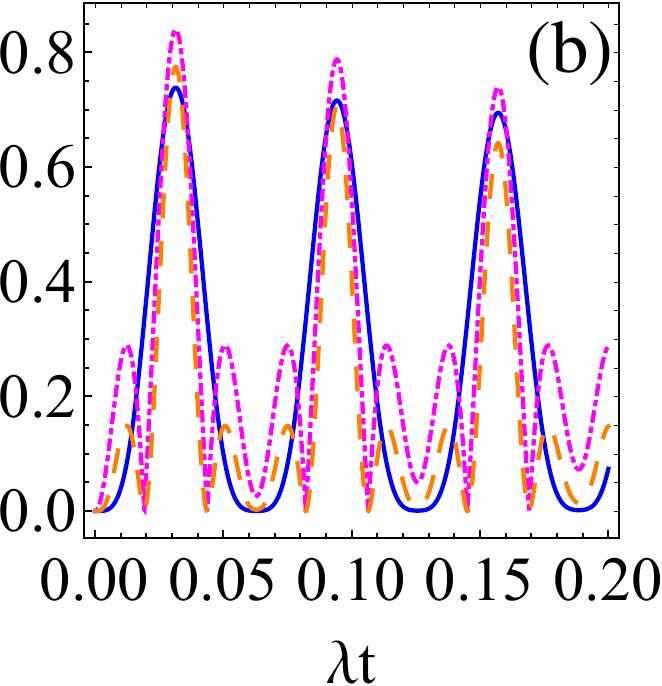}}
    
    \caption{(Color online) Dynamics in the Non-Markovian regime with charger-preferred coupling ($R=100\lambda$). (a) Revival of coherence: Unlike the rapid decay in Fig.~4, memory effects induce sustained oscillations in $D_{coh}$ (black dashed curve) and $C_{l_1}$ (green dotted curve). (b) Suboptimal energy recovery: Battery energy $E_B$ (blue solid curve) and entanglement (pink dot-dashed curve/orange long-dashed curve) are partially restored by backflow but remain lower than in the battery-preferred case (Fig.~3). The parameters are $\xi_1=\sqrt{3}/2, \xi_2=0.5$.}
    \label{f5}
\end{figure}


  \subsection{Non-Markovian Regime with Battery-Preferred Coupling}

We now turn to the strong coupling regime ($R=100\lambda$), where non-Markovian memory effects become dominant. We maintain the battery-preferred coupling configuration ($\xi_1 = 0.5$, $\xi_2 = \sqrt{3}/2$). The corresponding dynamics are illustrated in Fig. \ref{f3}.

As depicted in Fig. \ref{f3}(a), the first-order coherence $D_{coh}$ (black dashed curve) exhibits high-amplitude periodic oscillations without irreversible decay, in stark contrast to the monotonic decline observed in the Markovian limit. This behavior demonstrates that the strong coupling allows the environment to act as a memory buffer: lost phase information flows back into the system before it is irretrievably lost. The $l_1$-norm coherence $C_{l_1}$ (green dotted curve) oscillates in resonance with $D_{coh}$, maintaining high amplitudes throughout the evolution. This suggests that non-Markovian dynamics provide a mechanism for the \textit{dynamical protection} of quantum coherence, enabling the system to retain its quantum superposition character over long timescales.

The entanglement dynamics, presented in Fig. \ref{f3}(b), reveal a striking feature of the non-Markovian regime: both concurrence $E_{C}$ (pink dot-dashed curve) and entanglement of formation $E_{f}$ (orange long-dashed curve) oscillate strictly synchronously with the battery's stored energy $E_{B}$ (blue solid curve). This indicates that the generation and dissipation of entanglement are modulated into a resonant mode by the environmental backflow. Specifically, the peaks of energy storage coincide perfectly with the peaks of quantum resources. This implies a \textit{cooperative mechanism} where the oscillating quantum correlations directly drive the reversible energy exchange between the charger and the battery, preventing the energy saturation limit seen in the Markovian case.

 \subsection{Markovian Regime with Charger-Preferred Coupling}

To elucidate the impact of coupling asymmetry, we reverse the configuration to the charger-preferred case ($\xi_1=\sqrt{3}/2, \xi_2=0.5$) while keeping the system in the Markovian weak-coupling regime ($R=0.3\lambda$). The results are shown in Fig. \ref{f4}.

Comparing the coherence dynamics in Fig. \ref{f4}(a) with the battery-preferred case (Fig. \ref{f2}(a)), we observe a much faster decay. The first-order coherence $D_{coh}$ (black dashed curve) drops precipitously, and notably, the reconstruction of the $l_1$-norm coherence $C_{l_1}$ (green dotted curve) is significantly suppressed. This occurs because the charger, which carries the initial excitation, is now more strongly coupled to the dissipative bath. Consequently, the ``leakage'' of quantum information into the environment outpaces the establishment of internal system correlations.

The detrimental effect of this configuration is even more evident in the energy transfer dynamics shown in Fig. \ref{f4}(b). The entanglement measures ($E_C$ and $E_f$) appear only transiently with reduced amplitudes. Crucially, the battery stored energy $E_B$ (blue solid curve) remains negligible throughout the evolution. This confirms that strengthening the charger-environment coupling facilitates energy dissipation rather than wireless transfer. Physically, the charger acts as a ``leaky bucket'': its strong interaction with the bath dissipates the initial energy before it can effectively tunnel to the battery via the phonon-mediated channel.

\begin{figure}[htbp]
    \centering
    \subfigure{\includegraphics[width=4.2cm]{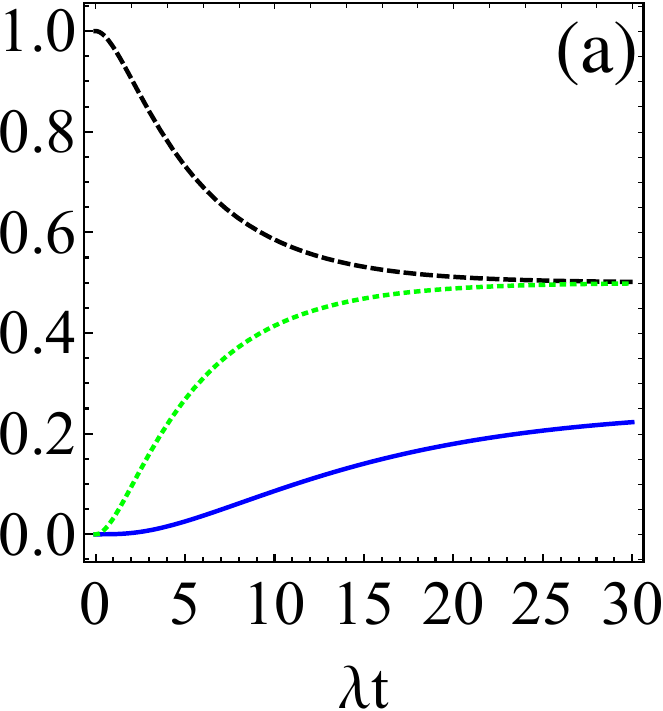}}
    \hspace{0.1cm}
    \subfigure{\includegraphics[width=4.2cm]{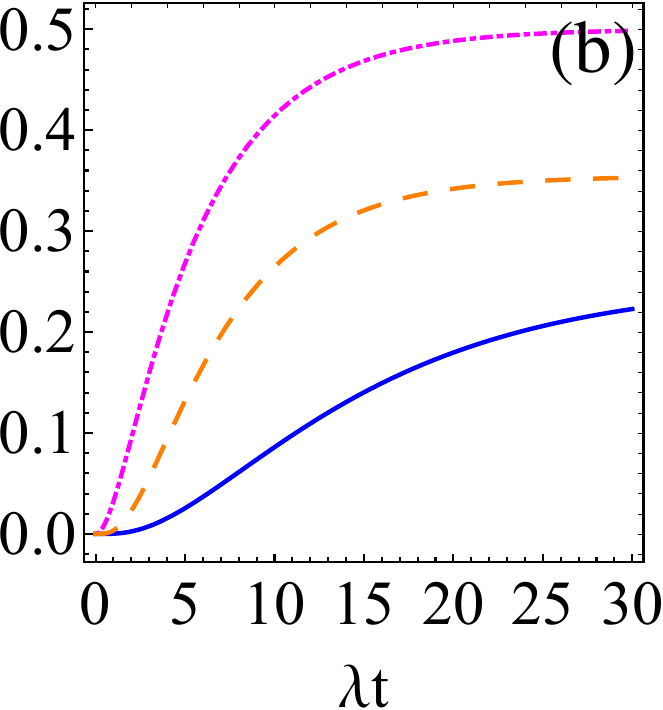}}
    
    \caption{(Color online) Dynamics in the Symmetric Coupling Markovian regime ($R=0.3\lambda$). (a) Resource consumption: Coherence $D_{coh}$ (black dashed curve) and $C_{l_1}$ (green dotted curve) decay monotonically. (b) Transformative dynamics: The accumulation of battery energy $E_B$ (blue solid curve) is accompanied by the depletion of entanglement $E_C/E_f$ (pink dot-dashed curve/orange long-dashed curve), indicating a conversion process from quantum correlations to stored energy. The parameters are $\xi_1=\xi_2=\sqrt{2}/2$ (symmetric coupling).}
    \label{f6}
\end{figure}

\begin{figure}[htbp]
    \centering
    \subfigure{\includegraphics[width=4.2cm]{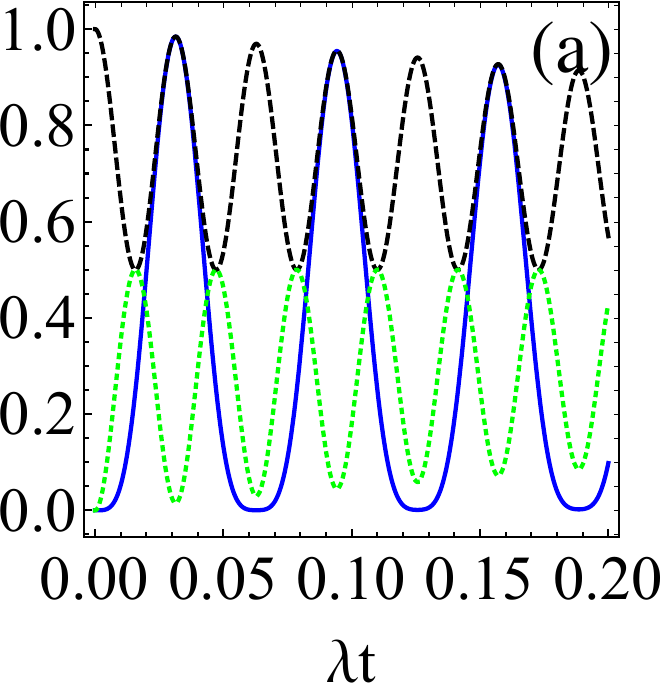}}
    \hspace{0.1cm}
    \subfigure{\includegraphics[width=4.2cm]{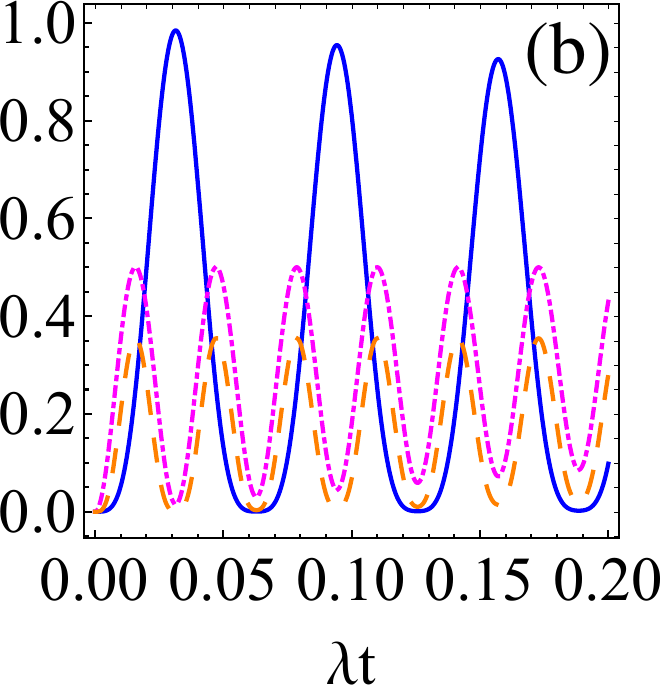}}
    
    \caption{(Color online) Dynamics in the Symmetric Coupling Non-Markovian regime ($R=100\lambda$). (a) Dark state protection: Coherence $D_{coh}$ (black dashed curve) and $C_{l_1}$ (green dotted curve) show persistent, high-amplitude oscillations due to decoupling from the environment. (b) Trapped energy dynamics: Battery energy $E_B$ (blue solid curve) and entanglement (pink dot-dashed curve/orange long-dashed curve) oscillate with a phase shift distinct from the asymmetric case, indicating energy trapping induced by the symmetric coupling. The parameters are $\xi_1=\xi_2=\sqrt{2}/2$.}
    \label{f7}
\end{figure}

\subsection{Non-Markovian Regime with Charger-Preferred Coupling}

Finally, we examine the charger-preferred configuration ($\xi_1=\sqrt{3}/2, \xi_2=0.5$) under the non-Markovian strong coupling condition ($R=100\lambda$). The results are depicted in Fig. \ref{f5}.

Comparing the dynamics in Fig. \ref{f5}(a) to the Markovian counterpart (Fig. \ref{f4}), the most prominent feature is the revival of quantum dynamics. The fast monotonic decay observed in the weak-coupling limit is replaced by sustained oscillations in both first-order coherence $D_{coh}$ (black dashed curve) and $l_1$-norm coherence $C_{l_1}$ (green dotted curve). This confirms that environmental memory effects effectively counteract the rapid dissipation caused by the strong charger-bath coupling, allowing quantum information to persist in the system despite the unfavorable coupling asymmetry.

While memory effects restore the oscillatory exchange of energy, the charging efficiency remains suboptimal compared to the battery-preferred case (Fig. \ref{f3}). As shown in Fig. \ref{f5}(b), although the battery energy $E_B$ (blue solid curve) oscillates and reaches non-zero values (unlike the negligible storage in Fig. \ref{f4}), its peak amplitude is suppressed. The entanglement measures ($E_C$ and $E_f$) also exhibit oscillatory behavior but with reduced maxima. This indicates that while non-Markovian backflow can mitigate the leaky bucket effect, the charger-dominated coupling still fundamentally hinders efficient energy localization at the battery site.

\subsection{Symmetric Coupling in Markovian Environment}

We now investigate the symmetric coupling configuration ($\xi_1 = \xi_2 = \sqrt{2}/2$) within the Markovian regime ($R=0.3\lambda$). This setup serves as a critical control group to understand the interplay between structural symmetry and resource dynamics. The results are presented in Fig. \ref{f6}}.

Unlike the asymmetric cases discussed previously, the symmetric configuration reveals a distinct transformative mechanism. As shown in Fig. \ref{f6}(a), the first-order coherence $D_{coh}$ (black dashed curve) exhibits a steady monotonic decay due to environmental dissipation. In contrast, the $l_{1}$-norm coherence $C_{l_{1}}$ (green dotted curve) gradually increases from zero and saturates, indicating that the system settles into a steady state rather than maintaining high-amplitude oscillations.

As depicted in Fig. \ref{f6}(b), in the Markovian regime with symmetric coupling, the entanglement dynamics exhibit a monotonic increase followed by saturation, notably avoiding the sudden death or oscillatory behaviors typically associated with non-Markovian memory effects.During the initial stage of energy accumulation, the rapid generation of entanglement establishes a robust quantum correlation channel between the subsystems. This rapid buildup actively drives the initial energy transfer from the charging field to the quantum battery. As time progresses, the entanglement curves plateau at stable, non-zero steady-state values. Physically, this indicates that the quantum transmission channel has reached its maximum capacity and remains stable. Therefore, under this symmetric coupling mechanism, entanglement is not consumed as a resource during the energy transfer process; rather, it serves as a persistent infrastructure that supports the continuous energy accumulation and maintains the subsequent steady state of the quantum battery.

\begin{figure}[htbp]
    \centering
 
    \subfigure{\includegraphics[width=4.2cm]{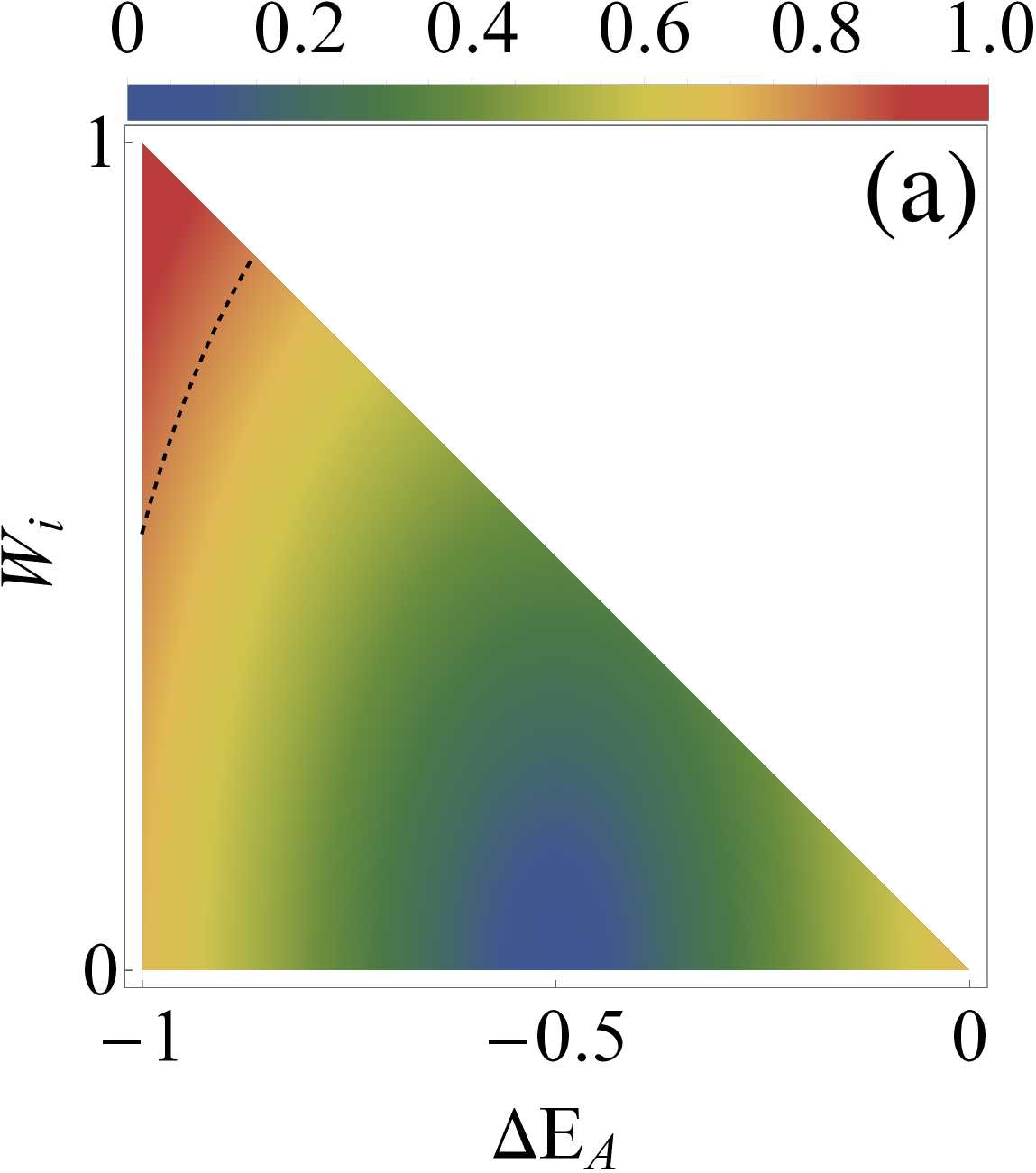}}
    \hspace{0.1cm}
    \subfigure{\includegraphics[width=4.0cm]{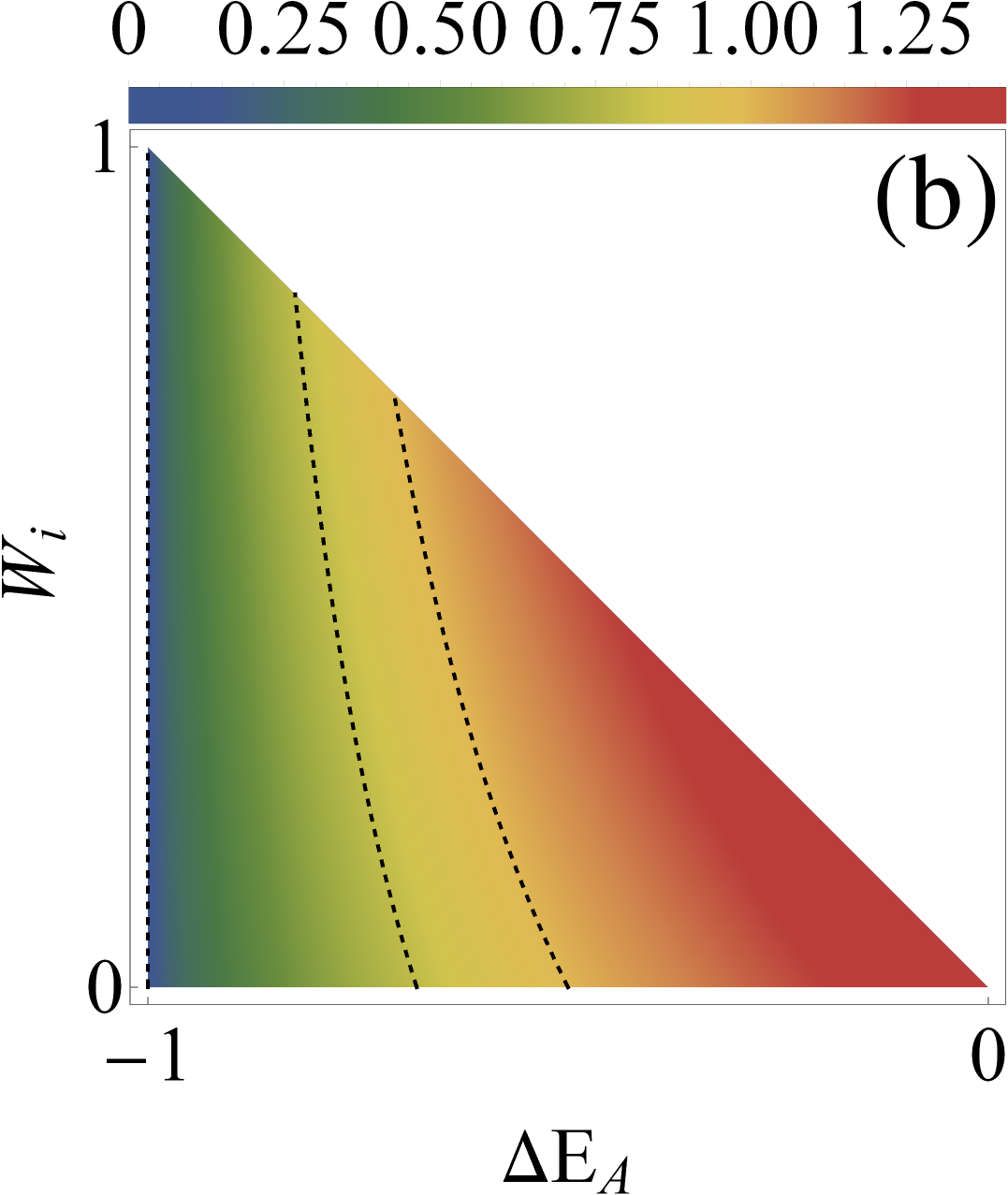}}
    
    \caption{(Color online) Thermodynamic trade-off between quantum resources and incoherent work cost. Density plots showing the distribution of (a) first-order coherence $D_{coh}$ and (b) $l_1$-norm coherence $C_{l_1}$ in the parameter space spanned by the charger's energy release $\Delta E_A$ ($x$-axis) and the incoherent work $W_i$ ($y$-axis). We set the initial state as $\left| {\Psi \left( 0 \right)} \right\rangle  = {\left| e \right\rangle _A}{\left| g \right\rangle _B}$, i.e., $\theta = \pi/2$ and $\alpha = 0$.}
    \label{f8}
\end{figure}

\subsection{Non-Markovian Regime with Symmetric Coupling}

Finally, we explore the distinct dynamics arising from the symmetric coupling ($\xi_1 = \xi_2 = \sqrt{2}/2$) in the non-Markovian strong-coupling regime ($R=100\lambda$). The results, displayed in Fig. \ref{f7}, reveal the emergence of a \textit{dark-state protection mechanism} \cite{five}.

In contrast to the decay seen in Markovian cases, the coherence dynamics in Fig. \ref{f7}(a) exhibit robust, undamped oscillations. This behavior is attributed to the formation of a sub-radiant dark state, which effectively decouples from the dissipative environment. Consequently, the quantum information is trapped within the system, shielded from environmental noise by the symmetry of the coupling configuration.

The energy transfer dynamics in Fig. \ref{f7}(b) further highlight the uniqueness of the symmetric regime. Unlike the perfect synchronization observed in the asymmetric cooperative mode (Fig. \ref{f3}), here the oscillations of entanglement ($E_C, E_f$) and battery energy ($E_B$) exhibit a distinct phase relationship. The energy exchange is driven by the coherent oscillation between the bright (super-radiant) and dark (sub-radiant) subspaces. This \textit{symmetry-protected trapping} ensures that a significant portion of energy and quantum resources remains accessible in the long-time limit, offering a strategy for stable quantum energy storage.

{Furthermore, after applying the BLP measure \cite{apdxa1} (quantum information backflow based on trace distance) for non-Markovianity in \hyperlink{A}{Appendix A}, we compare the non-Markovianity in different charging scenarios with the dynamic behavior of energy storage to reveal that the memory effect is a key factor in optimizing battery performance.}

\begin{figure*}
    \centering  
    \begin{minipage}{1.0\textwidth}
        \centering
        \subfigure{\includegraphics[height=4.3cm]{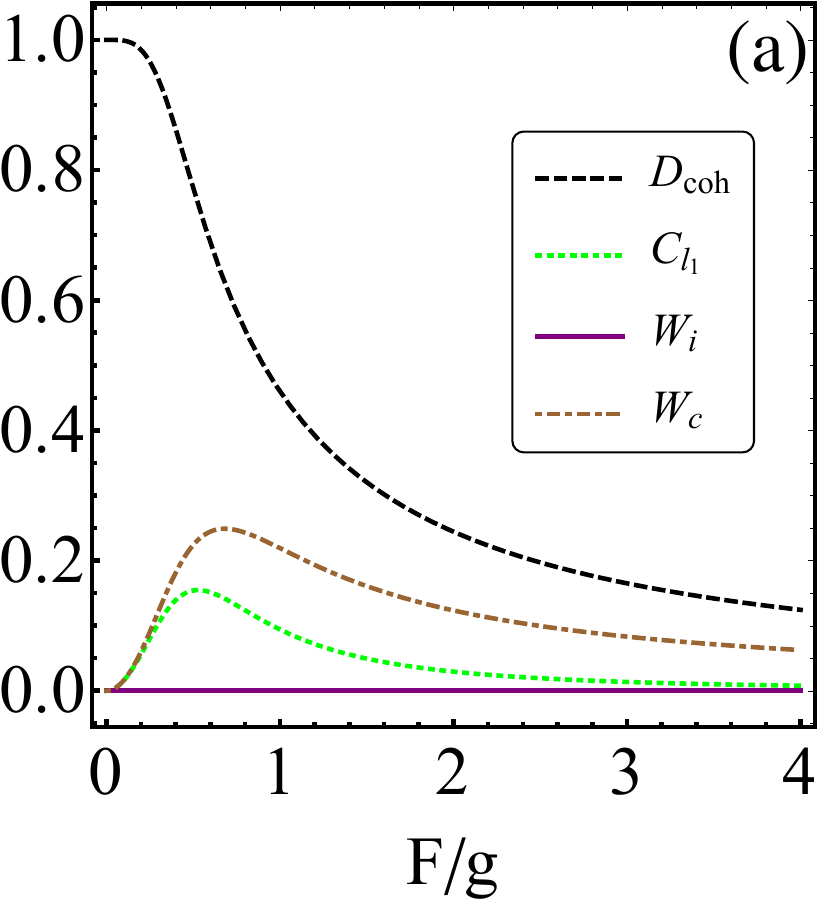}}
        \hfill 
        \subfigure{\includegraphics[height=4.3cm]{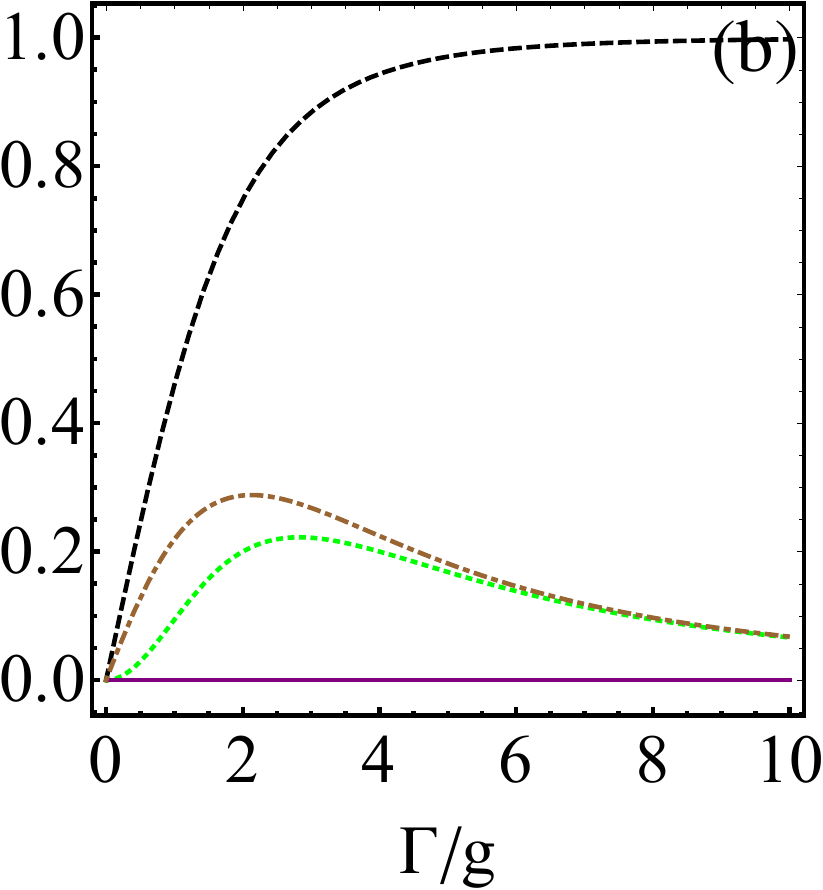}}
        \hfill
        \subfigure{\includegraphics[height=4.3cm]{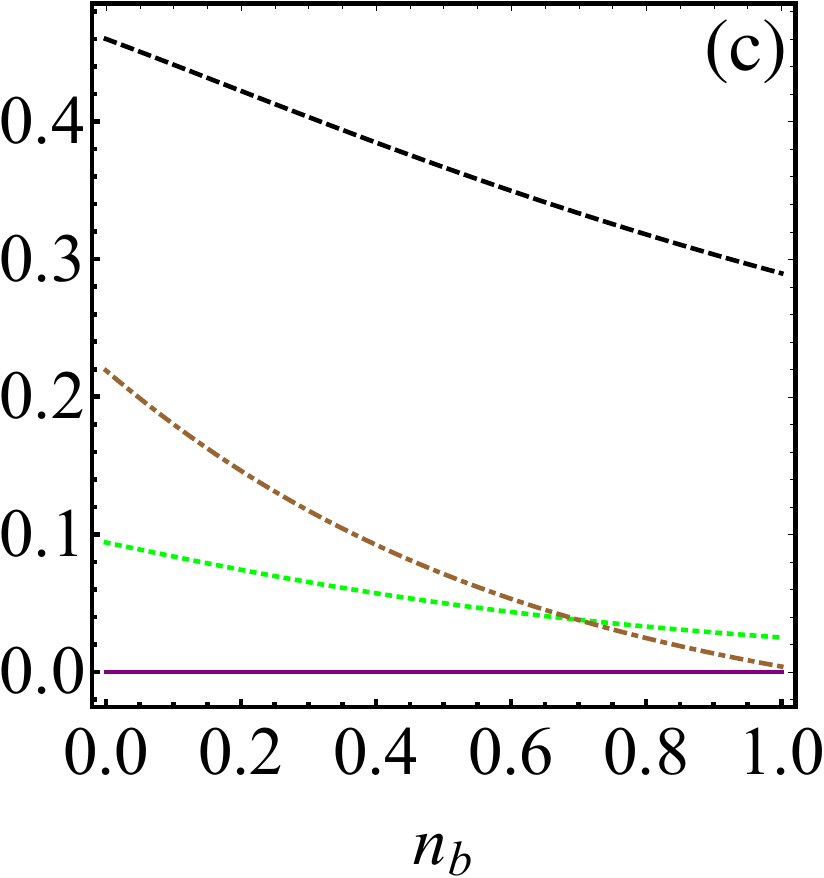}}
        \hfill
        \subfigure{\includegraphics[height=4.3cm]{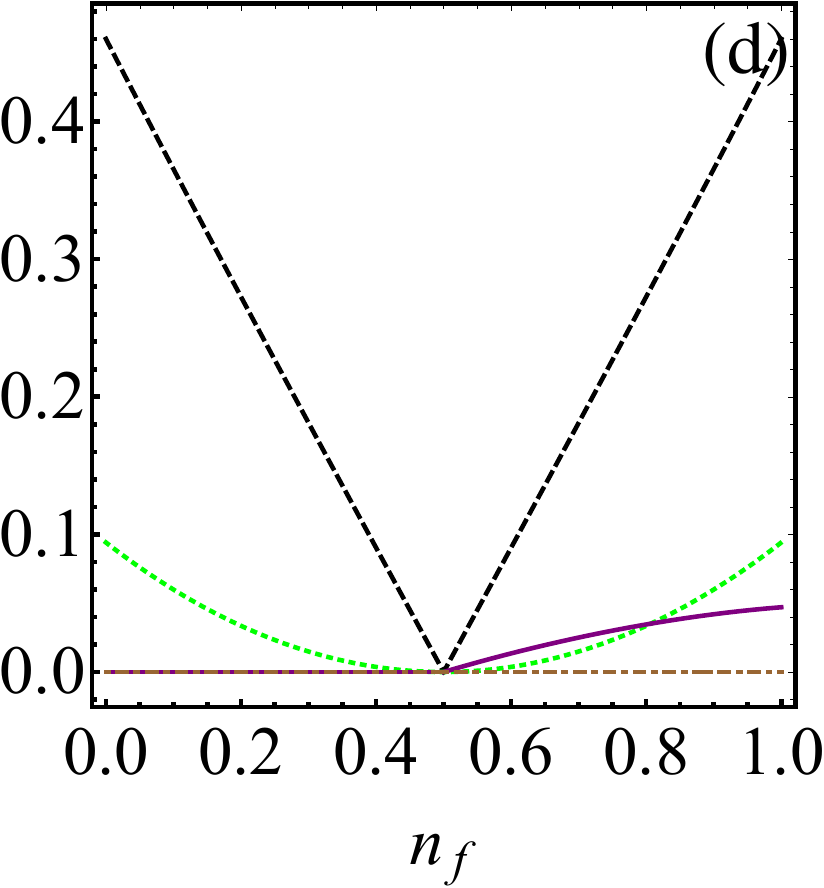}}
     
    \end{minipage}
    \caption{(Color online) The evolution of the relative entropy of coherence $D_{coh}$ (black dashed curve), the $l_1$-norm coherence $C_{l_1}$ (green dotted curve), the incoherent work $W_i$ (purple solid curve), and the coherent work $W_c$ (brown dot-dashed curve) are plotted under different conditions:
(a) As a function of the coupling strength $F/g$, with the dissipation rate fixed at $\Gamma = g$ and temperature $T = 0$.
(b) As a function of the dissipation rate $\Gamma/g$, with the coupling strength fixed at $F = g$ and temperature $T = 0$.
(c) As a function of the bath thermal occupation $n_b$, with $F=\Gamma = g$.
(d) As a function of the fermionic occupation factor $n_f$, with $F=\Gamma = g$.} 
\label{f9}
\end{figure*}

\section{Coherent stimulation enhances useful work}

To strictly quantify the thermodynamic quality of the energy stored in the quantum battery, we adopt the notion of ergotropy ($\mathcal{E}$), which represents the maximum amount of work that can be extracted from a quantum system via cyclic unitary operations. For a battery state $\rho(t)$ with Hamiltonian $H_B$, the total ergotropy is defined as the difference between its internal energy and the energy of its corresponding passive state: 
\begin{equation} \mathcal{W} \equiv \mathcal{E}(\rho) = \text{Tr}(\rho H_B) - \text{Tr}(\rho_P H_B), 
\end{equation}
where $\rho_P = \sum_n r_n |\epsilon_n\rangle\langle\epsilon_n|$ is the passive state associated with $\rho$. It is obtained by rearranging the eigenvalues $r_n$ of $\rho$ in a non-increasing order ($r_0 \ge r_1 \ge \dots$) with respect to the increasing energy eigenvalues $\epsilon_n$ of $H_B$ ($\epsilon_0 \le \epsilon_1 \le \dots$). The passive state represents a thermodynamically dead configuration from which no further work can be extracted unitarily.

Following the resource-theoretic framework established in \cite{ak1} and \cite{ak2}, the total extractable work $\mathcal{W}$ can be decomposed into two distinct contributions based on their physical origins: the incoherent work ($W_i$) and the coherent work ($W_c$).

The incoherent work $W_i$ quantifies the portion of work arising purely from population inversion (a classical non-equilibrium resource). It is formally defined as the ergotropy of the dephased state $\rho_{diag} = \Delta(\rho)$, where $\Delta$ is the dephasing map that removes all off-diagonal elements in the energy basis:

\begin{equation} 
W_i \equiv \mathcal{E}(\rho_{\mathrm{diag}}) = \text{Tr}(\rho_{\mathrm{diag}} H_B) - \text{Tr}(\rho_P H_B). 
\end{equation}

Since the dephasing operation $\Delta$ preserves the diagonal populations, the passive state of $\rho_{diag}$ is identical to $\rho_P$ . Thus, $W_i$ captures the work capacity that relies solely on the probability distribution of energy levels, disregarding any phase information.

The coherent work $W_c$ isolates the contribution specifically stemming from quantum coherence. It is defined as the difference between the total ergotropy and the incoherent part: 

\begin{equation} 
W_c \equiv \mathcal{W} - W_i = \text{Tr}(\rho H_B) - \text{Tr}(\rho_{\mathrm{diag}} H_B). 
\end{equation}
Physically, $W_c$ represents the quantum advantage in work extraction—the additional energy that becomes accessible only when the discharge protocol can coherently manipulate the superposition states.

\subsection{Single-unit battery}

To gain a deeper understanding of the thermodynamic quality of the energy transfer, we investigate the relationship between the energy released by the charger, the extractable work cost, and the available quantum resources. Assuming the initial charging system energy is concentrated within the charger, the ergotropy done by the battery consists entirely of incoherent components, written as

\begin{equation} 
\mathcal{W} = {W_i} = 2{\left| {{v_2}\left( t \right)} \right|^2} - 1.
\end{equation} 
{Meanwhile, the coherence and entanglement are interpreted as: \begin{align}
 & {D_{coh}} = {\left[ {2\Delta {E_A}\left( {\Delta {E_A} + 1} \right) + {{\left( {{W_i} + 1} \right)}^2}/2 - {W_i}} \right]^{1/2}}, \hfill \\
 & {C_{{l_1}}} = E_C = 2\left| {\sqrt {\left( {\Delta {E_A} + 1} \right)\left( {{W_i} + 1} \right)/2} } \right| ,\hfill \\  
& E_f = - \zeta \log_2 \zeta - (1 - \zeta) \log_2 (1 - \zeta),
		\label{Eq.dc}
	\end{align}
where, $\zeta = ({1 + \sqrt{1 - E_C^2}})/{2}$.}
Then we will discuss the potential for optimizing incoherent work through coherence in wireless protocol.
 
First-order coherence operates as a selective activation mechanism that governs the efficiency ceiling of incoherent work extraction. As depicted in Fig. \ref{f8}(a), high incoherent work regions ($W_i \rightarrow 1$) are strictly confined to zones exhibiting near-maximal $D_{\text{coh}}$ (approaching 1), with coherence exhibiting a precipitous decline along the dashed boundary as $W_i$ decreases. This pronounced gradient indicates that $D_{\text{coh}}$ establishes a stringent threshold for effective work extraction: substantial incoherent work can only be achieved when first-order coherence is maintained near its optimal value. Consequently, $D_{\text{coh}}$ acts as a precision bottleneck that determines the upper efficiency limit; once this coherence metric degrades, the enhancement of incoherent work becomes strictly constrained regardless of other parameters.

In contrast, $l_1$-norm coherence functions as a robust resource reservoir providing sustained foundational support for incoherent work generation. As illustrated in Fig. \ref{f8}(b), high $C_{l_1}$ values span a considerably broader parameter space, extending diagonally across the diagram with gentle transitions. Notably, $C_{l_1}$ maintains moderate levels even at intermediate $W_i$ values, exhibiting a gradual decay rather than a sharp cutoff. This extensive distribution demonstrates that $l_1$-norm coherence serves as the generic fuel for work extraction—operating without stringent conditional requirements and ensuring feasible work output across diverse operational regimes. It enables progressive enhancement and resilient performance of incoherent work across wide ranges of energy release and work output.
 
Collectively, the enhancement of incoherent work adheres to a threshold–fuel dual mechanism: First-order coherence $D_{\text{coh}}$ establishes the rigorous activation threshold required to unlock high-efficiency extraction (determining whether efficient work extraction is possible), whereas $l_1$-norm coherence $C_{l_1}$ constitutes the volumetric resource base that sustains work output under variable conditions (determining how much work can be extracted). The former operates as a precision switch mandating optimal coherence for peak performance, while the latter functions as a strategic energy repository ensuring operational robustness. Maximization of incoherent work necessitates both the fulfillment of the $D_{\text{coh}}$ threshold criterion and the availability of sufficient $C_{l_1}$ resources, thereby integrating stringent selectivity with resource abundance.

\subsection{Two-unit battery}
Next, we investigate the contribution of internal coherence within a two-unit battery to the two components of useful work. Specifically, we consider a system where a driven battery pack is immersed in a reservoir. In the interaction picture, the Hamiltonian of the battery pack is
\begin{equation} 
{H_B} = \sum\limits_{i = 1,2} {{\delta _i}\sigma _i^ + \sigma _i^ - }  + \left[ {g \sigma _1^ + \sigma _2^ -  + F\left( {\sigma _1^ +  + \sigma _2^ + } \right) + \rm{H.c.}} \right] ,
\label{eq18}
\end{equation} 
where $\delta _i=\omega_L-\omega_0$ represents the detuning between the battery pack with frequency $\omega_0$ and the driving external field with frequency $\omega_L$.  $\sigma _i^{ + \left(  -  \right)}$ denotes the Pauli raising (lowering) operator, while $g$ and $F$ represent the coupling strength and driving strength within the battery pack, respectively.
And the Lindblad equation for the charging system is
\begin{align}	
    {{\dot \rho }_B} &=  - i\left[ {{H_B},{\rho _B}} \right]  \nonumber\\
   &+ \Gamma \left( {N\left( T \right)\sum\limits_{i = 1,2} {{L_{\sigma _i^ - }}\left[ {{\rho _B}} \right]}  + n\left( T \right)\sum\limits_{i = 1,2} {{L_{\sigma _i^ + }}\left[ {{\rho _B}} \right]} } \right) 
		\label{Eq19}
	\end{align}
where ${L_\Lambda}\left[ {\rho}  \right] = 2\Lambda{\rho} {\Lambda^\dag } - {\Lambda^\dag }\Lambda \rho  - {\rho} {\Lambda^\dag }\Lambda$ denotes the dissipator, and $\Gamma$ is the dissipative rate. Regarding the different types of reservoirs, the average particle number $n(T)$ is different at frequency ${\omega_k }$ and temperature $T$. Specifically, $n\left( T  \right) = {n_b} = 1/\left[ {\exp \left( {\omega_k /T} \right) - 1} \right] $ and $N\left( T  \right) = 1 + n\left( T  \right)$ in the bosonic reservoir. For the fermionic one, $n\left( T  \right) = {n_f} = 1/\left[ {\exp \left( {\omega_k  - \mu } \right)/T + 1} \right]$ and $N\left( T  \right) = 1 - n\left( T  \right)$, where  $\mu $ denotes the chemical potential of the reservoir.

Fig. \ref{f9}(a) reveals the critical activation role of the coupling strength $F/g$. While the global coherence resource ($D_{coh}$) is maximal at the outset and decays monotonically with increasing interaction, it does not directly translate into extractable work. Instead, the extractable work ($W_c$) is strictly governed by the generation of $l_1$-norm coherence ($C_{l_1}$).
To enhance work extraction, one must tune the coupling to the sweet spot near $F/g \approx 0.6$. In this regime, the system effectively converts the decaying global coherence into useful local coherence ($C_{l_1}$), thereby maximizing $W_c$. Operating below this point fails to ignite the coherent mechanism, while operating above it leads to resource waste (decoherence) without work gain.

Fig. \ref{f9}(b) illustrates the destructive nature of environmental dissipation ($\Gamma/g$) on work extraction. As the decay rate increases, we observe a synchronized suppression of both the active coherence ($C_{l_1}$) and the coherent work ($W_c$). Unlike the coupling parameter $F/g$ which has an optimal peak, dissipation acts solely as a detriment.
To enhance work in the presence of dissipation, the strategy must focus on timescale optimization. Since $C_{l_1}$ is fragile against $\Gamma$, the work extraction process must be completed faster than the characteristic decay time defined by $\Gamma^{-1}$. The persistence of the correlation between $C_{l_1}$ and $W_c$ even under high dissipation confirms that protecting specific coherent basis states is more critical than preserving total system purity.

Fig. \ref{f9}(c) demonstrates the sensitivity of the quantum battery to thermal noise, quantified by the bath occupation number $n_b$. An increase in $n_b$ introduces thermal fluctuations that mix the quantum states, directly attacking the resource $C_{l_1}$.
To enhance work, the environment must be maintained at a low effective temperature (low $n_b$). The decline in $W_c$ with increasing $n_b$ reinforces the fact that thermalization is the enemy of coherent work. If $n_b$ cannot be reduced, one must employ dynamic decoupling or error-correction techniques to shield the specific $C_{l_1}$ coherence from the thermal bath, as the passive $D_{coh}$ alone cannot sustain work output against thermal noise.

As shown in Fig. \ref{f9}(d), (i) Monotonic enhancement of incoherent work:Incoherent work $W_i$ increases monotonically as the $n_f$ deviates from the equilibrium center ($n_f=0.5$), indicating that work extraction relies on breaking particle-hole symmetry to establish a chemical potential gradient. (ii) Complete suppression of coherent work: Coherent work $W_c$ remains strictly zero across the entire parameter space, demonstrating that quantum coherence cannot be converted into coherent work in the Fermionic reservoir-driven regime. (iii) Concomitant revival of quantum coherence: Both $D_{\text{coh}}$ (V-shaped) and $C_{l_1}$ (U-shaped) vanish at the equilibrium point but resurge symmetrically as $n_f$ departs from center. This reveals that coherence acts as an auxiliary transport condition—necessary for sustaining non-equilibrium particle flow—yet serves merely as a concomitant feature rather than the direct fuel for $W_i$ extraction.

{Finally, we have thoroughly investigated the impact of coherence between battery cells on energy storage across various charging scenarios (see \hyperlink{B}{Appendix B}) and established their correlation. It is worth noting that due to the symmetry of the charging configuration, each battery cell possesses an identical amount of stored energy (i.e., $\Delta E_A = \Delta E_B$); therefore, we have focused our attention on the energy of only one storage unit ($\Delta E_B$). Nevertheless, the optimization effect of coherence on the energy storage of the battery pack is, in fact, directly proportional to the number of cells.}

	\section{CONCLUSION AND OUTLOOK}

In this study, we have provided a resource-theoretic perspective on the charging dynamics of wireless quantum batteries, explicitly identifying quantum coherence and entanglement as the fundamental fuels driving energy accumulation. By analyzing the interplay between environmental memory effects and coupling configurations, we elucidated how these quantum resources actively regulate the efficiency and quality of energy storage.

First, we demonstrated that entanglement acts as the indispensable mediator for efficient power transfer in the non-Markovian strong-coupling regime. We observed a strict cooperative resonance where the oscillation of concurrence and entanglement of formation synchronizes perfectly with energy storage. This confirms that non-local correlations are not merely byproducts of the interaction but the active channel enabling reversible and rapid energy exchange between the charger and the battery. Crucially, we identified coupling configuration as a strategic switch for regulating these dynamics. We showed that asymmetric coupling (favoring the battery) significantly enhances charging speed by directing the energy flow preferentially toward the battery cell . In contrast, symmetric coupling in the non-Markovian regime unlocks a dark-state protection mechanism, effectively decoupling the system from the environment to lock the stored energy within a decoherence-free subspace .

Second, we revealed the critical compensatory role of quantum coherence in overcoming environmental dissipation. In the Markovian weak-coupling limit, we identified a mechanism where the system actively reconstructs $l_1$-norm coherence to compensate for the irreversible decay of first-order coherence . This reconstruction serves as a dynamic resource replenishment, sustaining the energy flow even when the initial phase information is degraded.

Most importantly, our thermodynamic analysis establishes a precise quantitative link between coherence resources and the extractable work (ergotropy). We unveiled a threshold-fuel dual mechanism: first-order coherence ($D_{coh}$) functions as a strict activation threshold that determines the feasibility of incoherent work extraction, while $l_1$-norm coherence ($C_{l_1}$) serves as the volumetric resource reservoir that directly governs the magnitude of coherent work . This distinction proves that maximizing specific forms of coherence is a prerequisite for enhancing the thermodynamic quality of the stored energy.

In summary, our results indicate that quantum resources are not passive features but controllable degrees of freedom. Through the engineering of coupling symmetry and non-Markovian backflow, these resources can be manipulated to optimize both the quantity and thermodynamic utility of energy in next-generation quantum batteries.

\begin{acknowledgments}
This work was supported by the National Natural Science Foundation of China (Grant nos. 12475009 and 12075001), Anhui Provincial University Scientific Research Major Project (Grant no.
2024AH040008), Anhui Province Science and Technology Innovation
Project (Grant no. 202423r06050004), Anhui Provincial Natural Science Foundation (Grant no. 2508085ZD001), and Anhui Provincial Department
of Industry and Information Technology (Grant no. JB24044).
\end{acknowledgments}
	
\appendix
{\hypertarget{A}{\section{Non-Markovianity and Energy}}}
	
	\begin{figure}
		\begin{minipage}{0.5\textwidth}
			\centering
		\subfigure{\includegraphics[width=4.2cm]{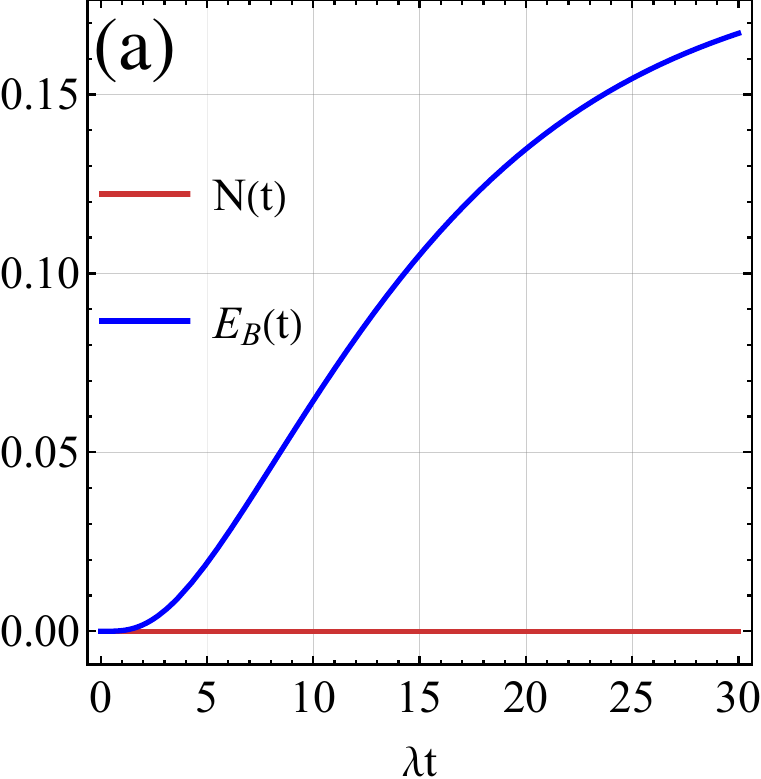}}
		\hspace{0.1cm}      
		\subfigure{\includegraphics[width=4.0cm]{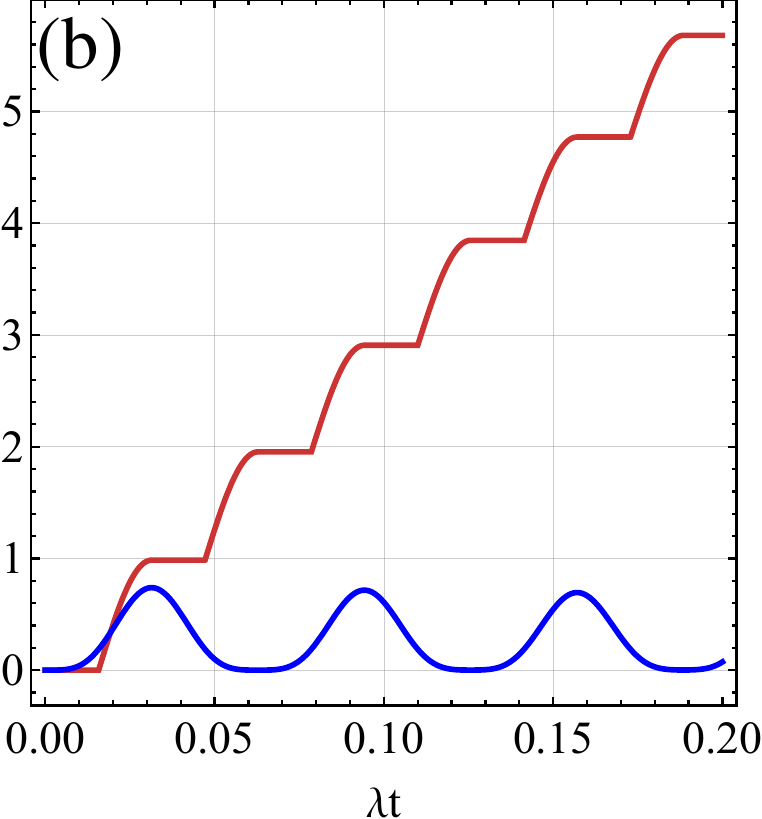}}
        \\
        \subfigure{\includegraphics[width=4.0cm]{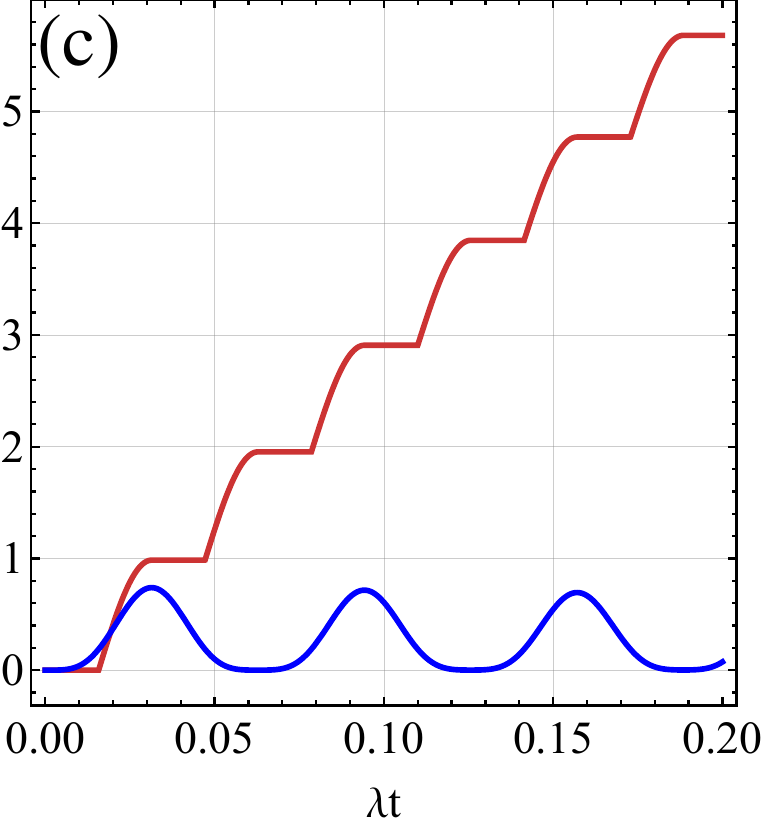}}
		\hspace{0.1cm}      
		\subfigure{\includegraphics[width=4.0cm]{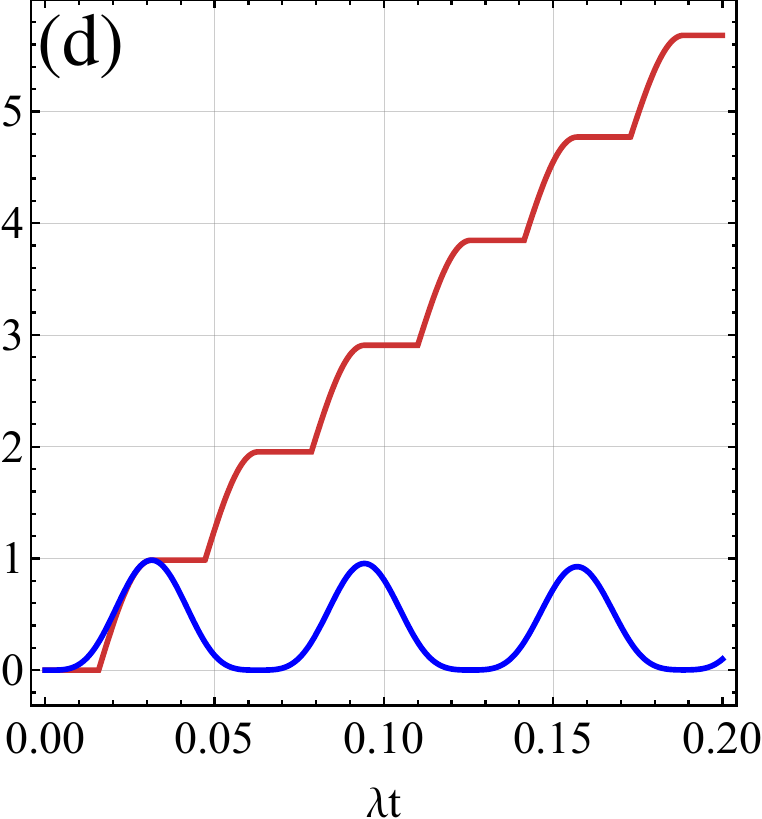}}
		\end{minipage}\hfill
		\caption{{ (Color online) Joint temporal evolution of the cumulative non-Markovianity $N(t)$ (wine-colored curves) and the battery energy $E_B(t)$ (blue curves).
(a) Markovian weak-coupling limit ($R=0.3\lambda$, $\xi_1=\sqrt{1}/2, \xi_2=\sqrt{3}/2$).
(b) Non-Markovian strong coupling (battery-preferred, $R=100\lambda$, $\xi_1=\sqrt{1}/2, \xi_2=\sqrt{3}/2$).
(c) Non-Markovian strong coupling (charger-preferred, $R=100\lambda$, $\xi_1=\sqrt{3}/2, \xi_2=\sqrt{1}/2$).
(d) Non-Markovian strong coupling (symmetric, $R=100\lambda$, $\xi_1=\xi_2=\sqrt{2}/2$).
        }}
		\label{f10}
	\end{figure}	

\begin{figure*}
    \centering  
    \begin{minipage}{1.0\textwidth}
        \centering
        \subfigure{\includegraphics[height=4.3cm]{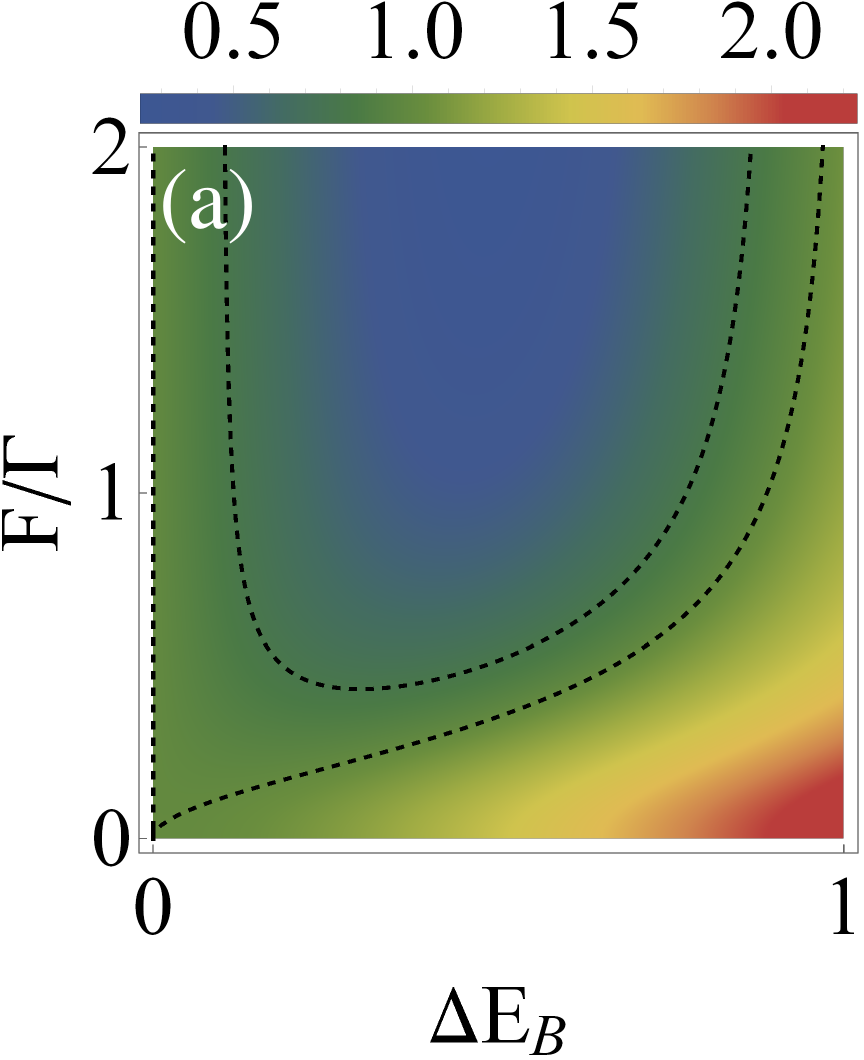}}
        \hspace{0.1cm} 
        \subfigure{\includegraphics[height=4.3cm]{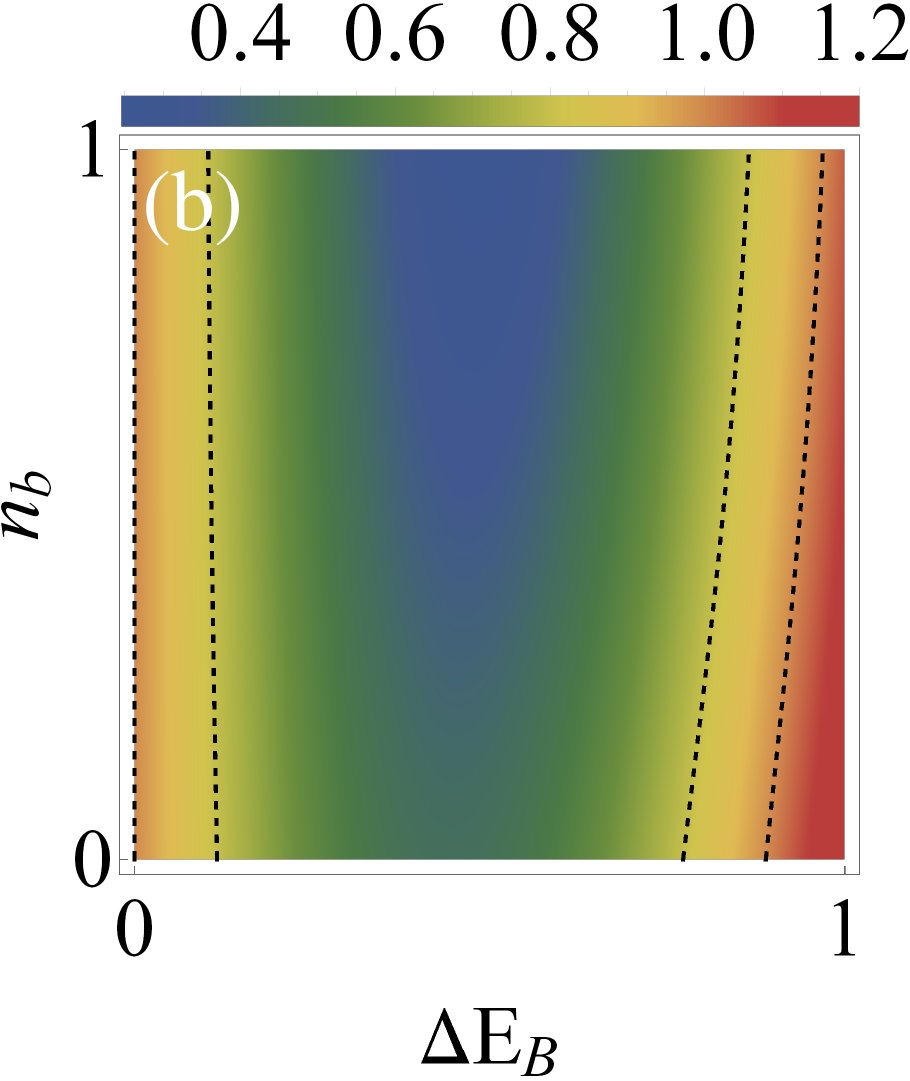}}
       \hspace{0.1cm}
        \subfigure{\includegraphics[height=4.3cm]{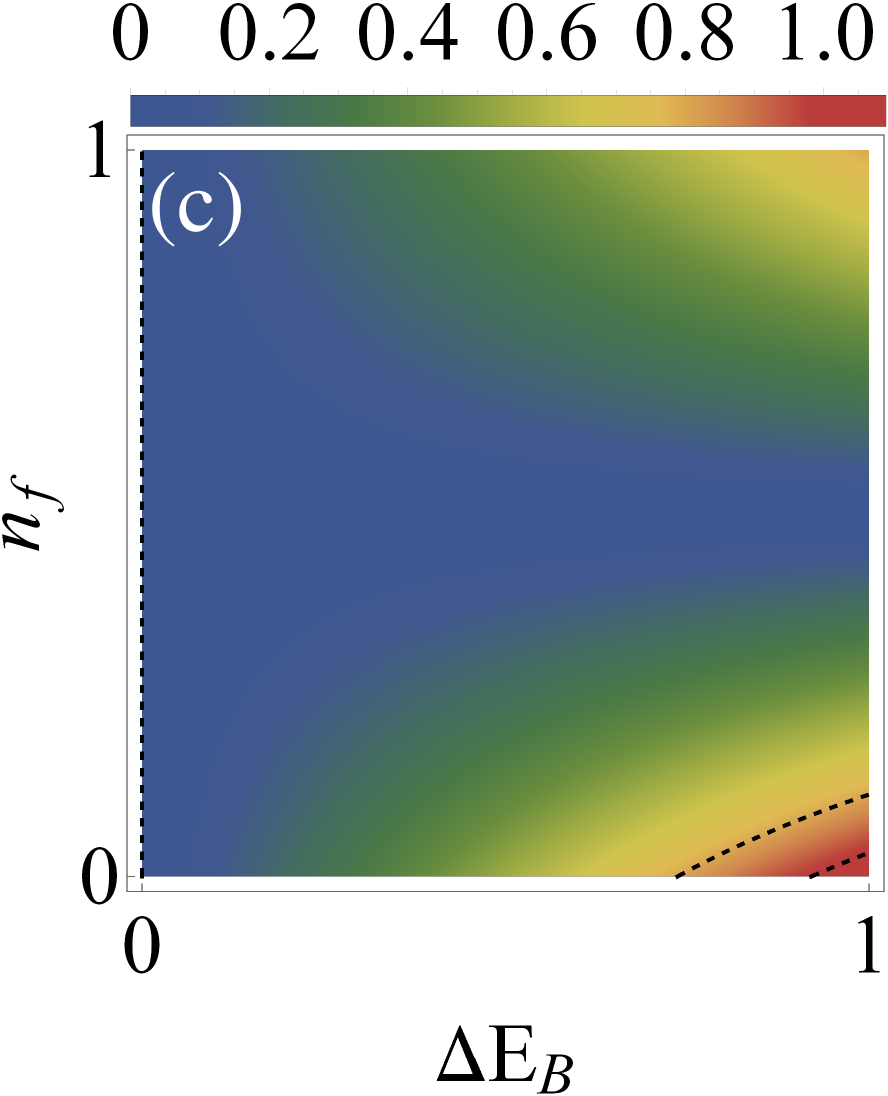}}
        \hspace{0.1cm}
        \\
         \subfigure{\includegraphics[height=4.3cm]{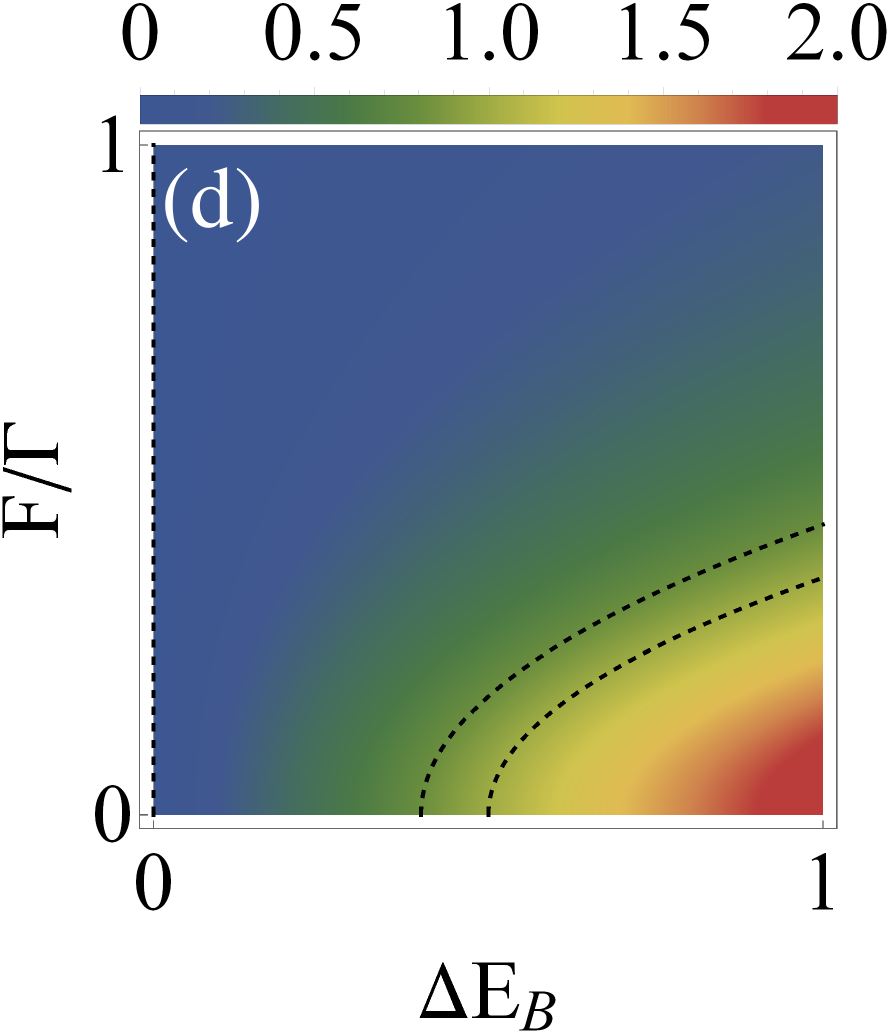}}
       \hspace{0.1cm} 
        \subfigure{\includegraphics[height=4.3cm]{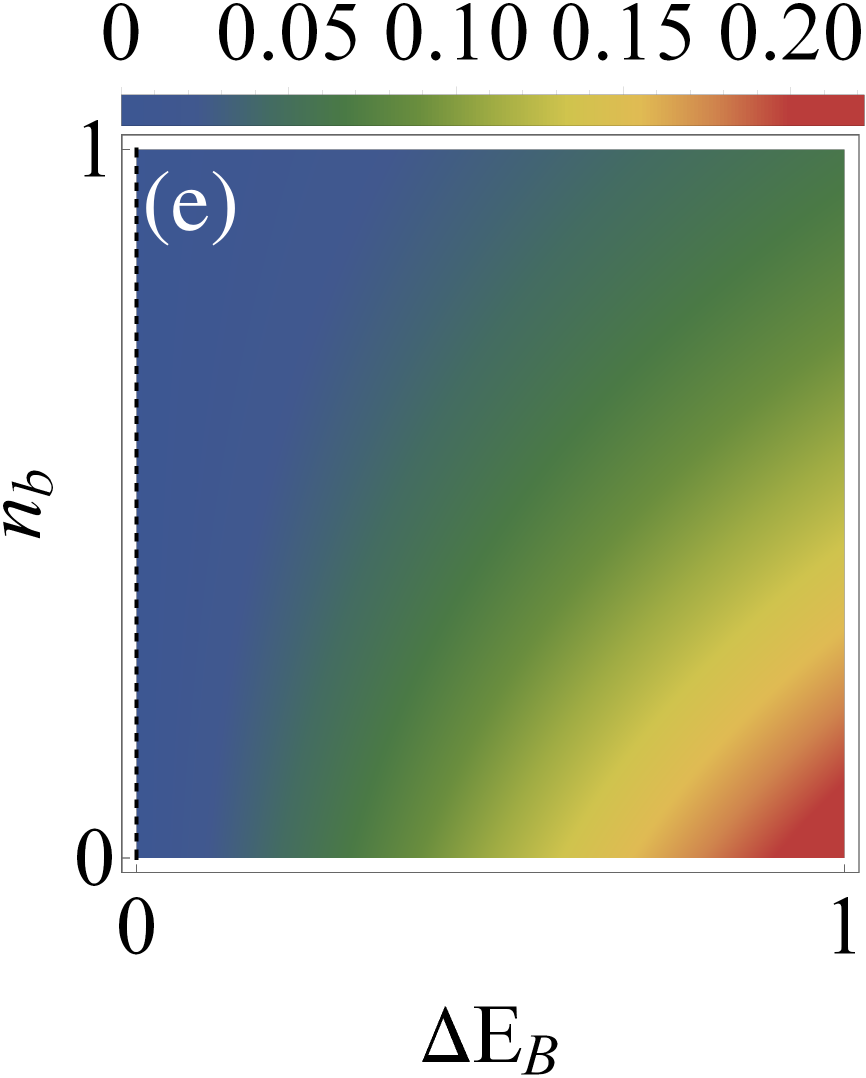}}
      \hspace{0.1cm}
        \subfigure{\includegraphics[height=4.3cm]{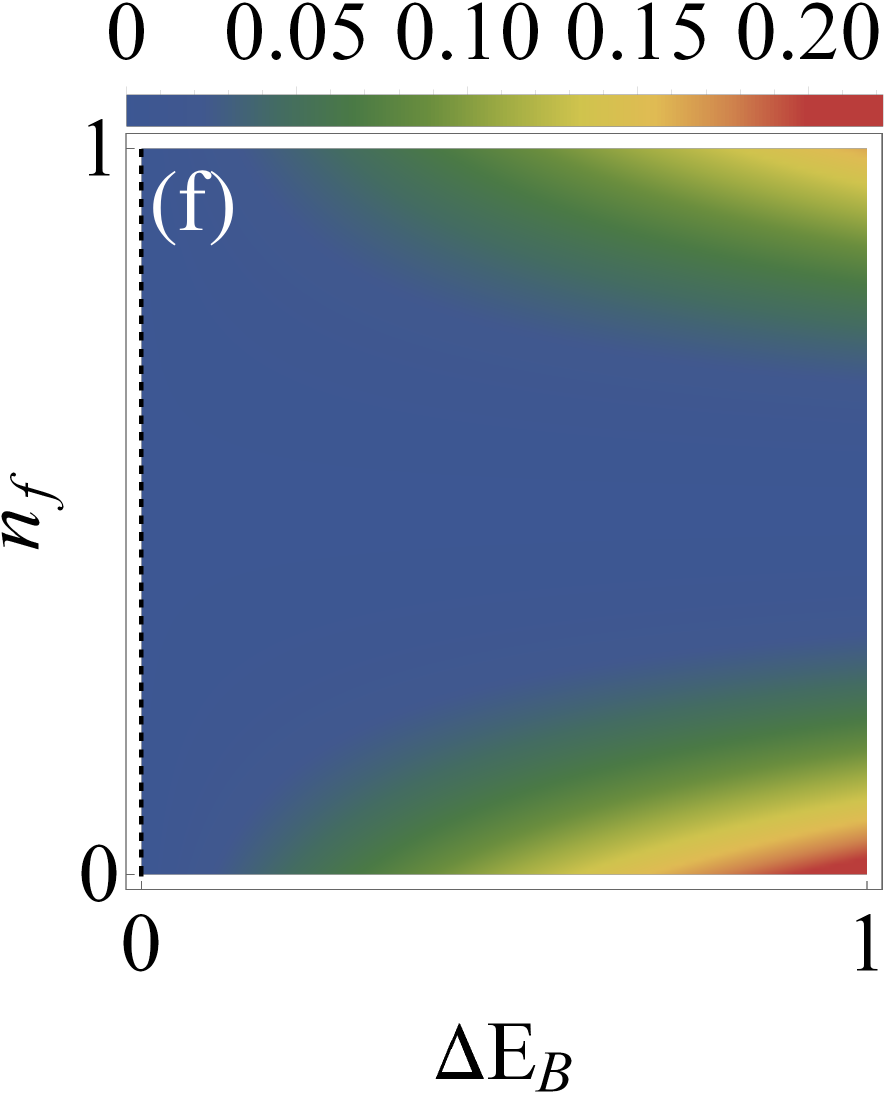}}
   
       \end{minipage}
    \caption{{(Color online) The landscape of first-order coherence $D_{coh}$ and $l_1$-norm coherence $C_{l_1}$ demonstrating the threshold-fuel dual mechanism under different environmental parameters as a function of the stored battery energy $\Delta E_B$.
(a)–(c) First-order coherence $D_{coh}$ plotted against $\Delta E_B$ under varying: (a) external driving strength $F/\Gamma$, (b) bosonic bath thermal occupation $n_b$, and (c) fermionic occupation factor $n_f$.
(d)–(f) $l_1$-norm coherence $C_{l_1}$ plotted against $\Delta E_B$ under varying: (d) external driving strength $F/\Gamma$, (e) bosonic bath thermal occupation $n_b$, and (f) fermionic occupation factor $n_f$.} }
\label{f11}
\end{figure*}

{In this Appendix, we conducted a detailed study on the impact of non-Markovianity on battery energy. Specifically, non-Markovianity is quantified as \cite{apdxa1}: 
$N_{\text{BLP}} = \max_{\rho_{1,2}(0)} \int_{\sigma > 0} \sigma(t, \rho_{1,2}(0)) \, dt$.
In the wireless charging model, we are focusing on the non-Markovianity, related to the decoherence function $\kappa(t)$: 
\begin{equation} 
N(t) = \int_{\frac{d}{dt}|\kappa(t)| > 0} \frac{d}{dt}|\kappa(t)| \, dt.
\end{equation} 
That is, non-Markovianity $N(t) $ is the integral of the trace distance derivative that is greater than zero over the time period.
}
    
{	
By introducing the quantitative non-Markovianity measure $N(t)$ and aligning it with the battery energy $E_B(t)$ over time, we clearly reveal how environmental memory drives wireless charging. In the Markovian weak-coupling limit ($R=0.3\lambda$ in Fig. \ref{f10}(a)), the non-Markovianity is strictly zero ($N(t) \equiv 0$). Without any information backflow, the shared environment acts as a one-way dissipation sink, restricting the battery to a painfully slow charge and a very low saturation capacity. Once the system is the non-Markovian strong-coupling regime ($R=100\lambda$ in Fig. \ref{f10}(b)-(d)), $N(t)$ exhibits a dramatic staircase-like growth. Physically, the environment now acts as a coherent pump, driving a robust backflow of escaping energy back into the system.}

{
Crucially, the ultimate storage efficiency of this backflow is strictly regulated by the coupling symmetry: In the battery-preferred configuration, this resource backflow synchronizes with the energy flow, triggering a high-power cooperative resonance (cf. Fig. \ref{f10}(b)). Under the charger-preferred setup, the non-Markovian backflow effectively suppresses complete dissipation (cf. Fig. \ref{f10}(c)), reviving the stored energy to a moderate level that would otherwise vanish in a memoryless bath.  The ultimate synergy occurs under completely symmetric coupling ($\xi_1=\xi_2$ in Fig. \ref{f10}(d)); here, while the non-Markovian backflow continuously retrieves coherent energy, the spatial symmetry creates a decoherence-free dark-state trap, securely locking the captured energy at its absolute full capacity ($E_B = 1.0$). These four sets of comparisons directly demonstrate that the charging power and maximum storage capacity of our wireless quantum battery are jointly governed by the quantitative non-Markovianity and coupling symmetry.
}

{\hypertarget{B}{\section{Threshold-Fuel Mechanism of Coherence}}}

	{In this appendix, we reveal the mechanism by which coherence operates on energy in complex charging protocols (cf. Eqs. (\ref{eq18}-\ref{Eq19})). When the resonant charging system is in a steady state, the coherence between batteries is related to the energy of the storage units, given by
\begin{widetext}
  $$
D_{\text{coh}} = \begin{cases} 
\sqrt{4{\Delta E_B}^2 - \frac{4\Delta E_B x}{x+1} + 1}, & T = 0 \\ 
\sqrt{(2\Delta E_B - 1)^2 + \frac{16B_1}{P}\Delta E_B}, & n = n_b, {F} = \Gamma = {g} \\ 
\frac{\left|\Delta E_B (1 - 2n_f)\right|\sqrt{B_2}}{\sqrt{1332 + n_f \left[609 + 64n_f(n_f - 3)\right]}}, & n = n_f, {F} = \Gamma = {g} 
\end{cases}, \ \   
C_{l_1} = \begin{cases}
\frac{2\Delta E_B}{x+1}, & T = 0 \\
\frac{8B_1}{P}\Delta E_B, & n = n_b, {F} = \Gamma = g \\
\frac{296\left|\Delta E_B\right|(1-2n_f)^2}{\left|1332 + n_f \left[609 + 64n_f(n_f - 3)\right]\right|}, & n = n_f, F = \Gamma = g
\end{cases}
$$
\end{widetext}
where, $x ={8F^2}/{\Gamma^2}$ , $B_1 = 37 + 36n_b(1 + n_b)$ ,$P = 1332 + n_b(2497 + 2n_b(2053 + (2n_b(123 + 2n_b(47 + 3n_b(5 + 2n_b))))))$, $B_2 = 2093201 + 640(n_f - 1)n_f (111 + 32(n_f - 1)n_f)$. 
}

    {To  elucidate the validity of the threshold-fuel dual mechanism under varying operational conditions, we inspect the behavior of first-order coherence ($D_{coh}$) and $l_1$-norm coherence ($C_{l_1}$) as a function of the accumulated battery energy ($\Delta E_B$) across several typical environmental profiles, as illustrated in Fig. \ref{f11}(a)–(c) depict $D_{coh}$ under the variation of the ratio of drive to dissipation ($F/\Gamma$), bath thermal occupation ($n_b$), and fermionic occupation factor ($n_f$). It is prominent that high battery energy accumulation ($\Delta E_B$) is strictly confined to a narrow, steep landscape where $D_{coh}$ remains near its maximum value. As the environmental parameters degrade the phase synchronization, $D_{coh}$ exhibits a precipitous, sharp-cutoff transition. This all-or-nothing boundary mathematically identifies $D_{coh}$ as the absolute activation threshold; without satisfying this rigid boundary condition, energy storage cannot be effectively triggered. Conversely, Fig. \ref{f11}(d)–(f) present the corresponding distribution of the $l_1$-norm coherence ($C_{l_1}$). In sharp contrast to the abrupt cutoff of $D_{coh}$, the landscape of $C_{l_1}$ displays a smooth, continuous, and wide-ranging gradient spanning across the $\Delta E_B$ axis. As the battery energy shifts, the volumetric resource capacity of $C_{l_1}$ adapts progressively, serving as a resilient energy repository. This smooth scaling fundamentally substantiates that $l_1$-norm coherence acts as the explicit, quantifiable fuel that determines the exact capacity and sustainability of the coherent charging process under diverse driving and dissipative environments.}

\end{document}